\documentclass[twocolumn,nopacs,preprintnumbers,amsmath,amssymb,aps,prb,reprint]{revtex4}
\usepackage{graphicx}
\begin{document}

\title{
Nonequilibrium Phases and Segregation for Skyrmions on Periodic Pinning Arrays   
} 
\author{
C. Reichhardt, D. Ray, and C. J. O. Reichhardt 
} 
\affiliation{
Theoretical Division and Center for Nonlinear Studies,
Los Alamos National Laboratory, Los Alamos, New Mexico 87545, USA\\ 
} 

\date{\today}
\begin{abstract}
Using particle-based simulations,
we examine the collective dynamics of skyrmions interacting with periodic pinning
arrays, focusing on the impact of the Magnus force on the sliding phases.  
As a function of increasing pinning strength,
we find a series of distinct dynamical phases,
including an interstitial flow phase, a 
moving disordered state, a moving crystal, 
and a  segregated cluster state.
The transitions between these states produce signatures in
the skyrmion lattice structure, the skyrmion Hall angle,
the velocity fluctuation distributions, and the velocity-force curves.
The moving clustered state is similar to the segregated state recently observed
in continuum-based simulations with strong quenched disorder.
The segregation arises from the drive dependence
of the skyrmion Hall angle,
and appears in the strong pinning limit when
the skyrmions have nonuniform velocities, causing
different portions of the sample to have different effective
skyrmion Hall angles. 
We map the evolution of the dynamic phases
as a function of the system density, the ratio of the Magnus force to
the dissipative term, and the ratio of the number 
of skyrmions to the number of pinning sites.
\end{abstract}
\maketitle

\section{Introduction}
Numerous systems can be modeled as 
a collection of interacting particles that are
coupled to an underlying periodic substrate. 
The ratio of the number of particles
to the number of substrate minima, known as the filling factor,
can be an integer, a rational fraction, or an irrational fraction.
As a function of 
filling factor,
different types of crystalline, partially disordered, and disordered states can arise 
depending on the symmetry of the 
underlying substrate \cite{8,9}. 
Commensuration effects arise for certain integer or rational filling
factors that
strongly affect the static and dynamic phases
\cite{1,2}.
Examples of systems exhibiting commensuration behavior
include atoms on ordered substrates \cite{3}, 
colloids interacting with
two-dimensional periodic substrates \cite{4,5,6,7,8},
vortices in type-II superconductors
with periodic pinning arrays \cite{9a,10a,9,10,11,12},
and vortices in Bose Einstein condensates
with optical trap arrays \cite{13,14}.
Under an applied 
drive, a rich variety of distinct sliding phases appear
\cite{2}. 
The critical driving force $F_c$ needed to induce depinning
typically exhibits a maximum at commensurate fillings,
while incommensurate fillings contain
localized regions  with a lower depinning threshold
that produce a flow of soliton-like excitations at depinning,
followed at increasing driving forces
by additional dynamical phases as the other particles depin
in a multi-step depinning process
\cite{2,7,8,9,15,16,17,18,19,20,21,22,23}. 
These systems also exhibit
a variety of different types of sliding phases,
including the soliton-like motion of interstitials or vacancies, grain boundary 
flow, chains of flowing particles moving around pinned particles, disordered
or strongly fluctuating flow phases,
and coherent moving crystal phases.
Transitions between the sliding phases
can be characterized by changes in the
velocity-force curves, fluctuations, and structure of the particles \cite{15}. 

Vortices in type-II superconductors pass through
a particularly rich set of static and dynamic phases when 
interacting with a periodic substrate.
Commensuration
effects produce
peaks in the critical current or the force needed to depin vortices 
at certain integer \cite{7,9,10,11,12,24} and
rational filling factors \cite{25,26,27}.
The pinning lattice
consists of localized sites that can capture one or more vortices.
Additional vortices that are not in the pinning sites
sit in the interstitial regions between 
occupied pinning sites,
and the combination of directly pinned vortices and
indirectly pinned interstitial vortices produces the
variety of dynamical phases
found in simulations \cite{16,25,28,29,30,31,32,33} and experiments \cite{8,24,34,35}.

Skyrmions in chiral magnets
\cite{36,37,38,39,40,41} have
many
similarities to superconducting vortices.
The skyrmions are particle-like magnetic
textures that form triangular lattices, can be driven with 
an applied current,
and also exhibit depinning and sliding phases \cite{38,42,43,44,45,46,47,48}.
One of the key differences between
skyrmions and superconducting vortices
is that the skyrmion dynamics is strongly influenced by
a Magnus force, produced by the skyrmion topology,
which generates
velocity components perpendicular to
the forces experienced by the skyrmion \cite{38,42,44,45}.
Due to the Magnus force, the
skyrmions move at an angle called the skyrmion Hall angle $\theta_{sk}$
with respect to the direction of the externally applied drive.
As a skyrmion passes through a pinning site, its trajectory describes an arc which
reduces the effective pinning threshold and also shifts the position of the skyrmion
in a side jump effect.  As a result, the skyrmion Hall angle
varies with applied current.
In the presence of pinning,
$\theta_{sk}=0$ at the depinning threshold,
and as the skyrmion velocity increases with increasing current,
$\theta_{sk}$ also increases
until it saturates to the clean or intrinsic value
$\theta^{\rm int}_{sk}$ at high drives.
This behavior
was first observed
in particle-based simulations \cite{49,50,51,52}
and was subsequently
verified in experiments \cite{53,54,55}.
Continuum-based simulations
with pinning  also show that the skyrmion Hall angle is drive dependent
\cite{56,57}, which provides further evidence that particle-based
simulations capture many of the essential features of the skyrmion dynamics.  

With the recent advances in creating nanostructured materials that support
skyrmions, it should be possible to create carefully controlled
periodic structures or pinning arrays for skyrmions,
and in this work we use particle-based simulations 
to investigate the dynamics of skyrmions in square pinning arrays.
In addition to the skyrmion physics,
we address the general question of how the
collective dynamics of particle assemblies on periodic substrates change 
in the presence of a non-dissipative Magnus force,
and what new dynamic phases arise that are absent in the overdamped limit.
Previous work on individual
skyrmions
moving over two-dimensional periodic substrates
showed that the Magnus force strongly affects the dynamics \cite{50}.
Specifically, the drive dependence of the skyrmion Hall angle 
produces quantized steps
in $\theta_{sk}$ as a function of driving current
when the skyrmion locks to different symmetry directions
of the substrate \cite{50}.

In this work we study
collectively interacting 
skyrmions on a periodic substrate, where new dynamics emerge that
are not present for an isolated skyrmion.
In the overdamped limit, we find interstitial flow, disordered flow, and
moving crystal flow phases, in agreement with previous work on
vortices in periodic pinning arrays
\cite{15,16,30,34,35}.
Upon introducing a nonzero Magnus force, these same three phases persist in the form
of
an interstitial flow that can lock to various symmetry angles of the
substrate, a disordered flow phase, and a moving crystal
phase.
When both the substrate strength and the Magnus force
are
sufficiently large, we observe a transition 
to a moving clustered or segregated state that is 
similar to the skyrmion cluster state
found in continuum-based simulations 
with strong pinning \cite{58}.
In our case, the clustering
instability arises from the strong velocity dependence of $\theta_{sk}$
that appears when the pinning is strong.
The skyrmions form bands
traveling at different relative velocities,
and these bands move toward one another.  
We observe a similar clustering
effect for strong randomly placed pinning which will be described elsewhere.
When the pinning is weak,
the clustering effect is lost and we
observe an Aubry type transition \cite{59,60} on the square pinning 
lattice, where the pinned skyrmion lattice transitions
to a floating triangular skyrmion lattice
with a much lower depinning threshold.
As the filling factor changes, we find
several commensurate-incommensurate transitions
leading to the stabilization of different types of pinned skyrmion crystalline states 
that are identical to
those that occur
in the overdamped limit since the ground configurations
are not affected by the Magnus force. 
We also show that it is possible to have different types of
moving segregated states such as a
diagonal stripe phase
in which the stripes move at an angle with respect to the
major symmetry axis of the pinning substrate.

The paper is organized as follows.  In Section II, we describe our simulation.
In Section III we illustrate the dynamic phases with emphasis on the
phase separated states, and we construct a dynamical
phase diagram as a function of drive and pinning strength.
Directional locking effects occur due to the periodicity of the pinning.
In Section IV we study
the effect of changing the relative strength of the
Magnus force on the dynamic phases.
The effect of varying the filling factor appears in Section V.
In Section VI we consider the effect of changing the pinning density 
while holding the filling factor constant.
We summarize our results in Section VII.

\section{Simulation}
We utilize a particle-based model for skyrmions
based on
a modified Thiele equation that takes into account skyrmion-skyrmion and
skyrmion-pin interactions \cite{45,49,50,51,52,62}.  
This model has previously been shown to capture several features of skyrmion
dynamics including the orientation of skyrmion lattices, pinning phenomena,
and the  drive dependence of the
skyrmion Hall angle
when pinning is present.   
We consider a two-dimensional system with periodic boundary conditions in
the $x$- and $y$-directions containing
$N_{sk}$ skyrmions and $N_{p}$ pinning sites.
The filling factor is $f = N_{sk}/N_{p}$. For most 
of this work we
hold the number of pinning sites fixed and vary $N_{sk}$, which can be achieved
experimentally
by changing the applied magnetic field.  

The skyrmion dynamics are governed by the following equation of motion
\begin{equation}
\alpha_d {\bf v}_{i} + \alpha_m {\hat z} \times {\bf v}_{i} =
{\bf F}^{ss}_{i} + {\bf F}^{p}_{i} + {\bf F}^{D}  .
\end{equation}
The velocity of skyrmion $i$ is
${\bf v}_{i} = {d {\bf r}_{i}}/{dt}$, while
$\alpha_d$  
and $\alpha_m$ are the damping and Magnus terms,
respectively.
The driving force
${\bf F}^{D}=F_D{\bf \hat{x}}$ represents an applied current.
In the absence of pinning the skyrmions move at the intrinsic skyrmion Hall angle
$\theta^{\rm int}_{sk} = \tan^{-1}(\alpha_{m}/\alpha_{d})$
with respect to the driving direction.
The skyrmion-skyrmion interaction force is repulsive and takes the form
${\bf F}_{i}^{ss} = \sum^{N_{sk}}_{j=1}K_{1}(r_{ij}){\hat {\bf r}_{ij}}$, 
where $K_{1}$ is the modified Bessel function, $r_{ij} = |{\bf r}_{i} - {\bf r}_{j}|$ 
is the distance between skyrmions $i$ and $j$,
and ${\hat {\bf r}}_{ij}=({\bf r}_i-{\bf r}_j)/r_{ij}$.
The pinning sites are modeled as
localized parabolic potentials with a fixed radius $R_{p}$ and strength $F_{p}$,
${\bf F}_i^{p} = \sum_{k=1}^{N_p}(F_{p}r_{ik}^p/R_{p})\Theta(R_{p} -r_{ik}^p)\hat{\bf r}_{ik}^p$
where $r_{ik}^{p}=|{\bf r}_{i}-{\bf r}_k^{p}|$,
$\hat {\bf r}_{ik}^p=({\bf r}_{i}-{\bf r}_k^{p})/r_{ik}^p$, and $\Theta$ is
the Heaviside step function.
This same pinning model was previously used
to study the statics and dynamics
of superconducting vortices in periodic pinning arrays \cite{16,25,28,30,31,32}.  

The initial skyrmion positions
are obtained by
simulated annealing,
in which we start from a high temperature liquid state and gradually
cool to $T=0$. 
After annealing we gradually increase
the driving force in increments of $\delta F_{D} = 0.002$,
waiting $20000$ simulation time steps between increments to ensure that the
system has reached a steady state.
For small values of $F_{p}$
we have tested increments as small as
$\delta F_D=1 \times 10^{-6}$ to verify that the results are not affected
by the choice of $\delta F_D$.
We measure $\langle V_{||}\rangle = N_{sk}^{-1}\sum^{N_{sk}}_{i=1} v_{||}^{i}$
and $\langle V_{\perp}\rangle = N_{sk}^{-1}\sum^{N_{sk}}_{i=1} v_{\perp}^{i}$
in the directions parallel and perpendicular to the drive, respectively,
where $v^i_{||}={\bf v}_i \cdot {\hat {\bf x}}$
and $v^{i}_{\perp}={\bf v}_i \cdot {\hat {\bf y}}$,
and we use these quantities to obtain
the velocity ratio $R=\langle V_{\perp}\rangle/\langle V_{||}\rangle$ and
the measured skyrmion Hall angle
$\theta_{sk} = \tan^{-1}(R)$.
We also measure
a quantity related to the standard deviation of the skyrmion velocities in the parallel
and perpendicular directions,
$\delta V_{||} = \sqrt{ [\sum_{i}^{N_{sk}} (v_{||}^i)^2 - \langle V_{||}\rangle^2]/N_{sk}}$  and
$\delta V_{\perp} = \sqrt{ [\sum_{i}^{N_{sk}} (v_{\perp}^i)^2 -\langle V_{\perp}\rangle^2]/N_{sk}}$.
To quantify the amount of ordering in the skyrmion lattice, we use the
fraction of sixfold-coordinated skyrmions, $P_6=N_{sk}^{-1}\sum_{i}^{N_{sk}}\delta(6-z_i)$,
where $z_i$ is the coordination number of skyrmion $i$ obtained from a Voronoi
tessellation.  For a perfect triangular lattice, $P_6=1$.
We fix $L = 36$, $R_{p} = 0.35$, and $N_{p} = 256$ while varying
$F_{p}$,  $\alpha_{m}/\alpha_{d}$, and $N_{sk}$, except in Section VI
where we
consider varied pinning density $n_p=N_{p}/L^2$.

\section{Dynamic Phases and Clustering}

\begin{figure}
\includegraphics[width=\columnwidth]{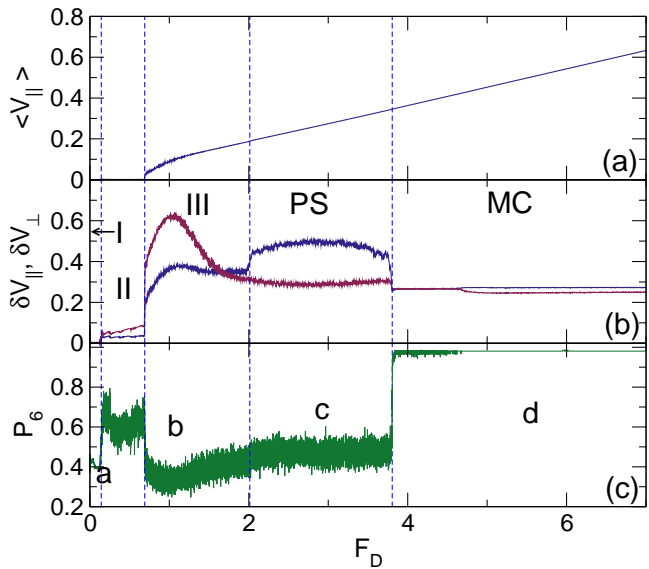}
\caption{(a) $\langle V_{||}\rangle$ vs $F_{D}$
for skyrmions moving in a square periodic pinning array
with a filling factor of $f=N_{sk}/N_p=1.0117$, pinning strength
$F_p=2.0$ and pinning density $n_p=0.1975$ at 
$\alpha_{m}/\alpha_{d} = 9.96$. 
(b) The corresponding velocity deviations
$\delta V_{||}$ (blue) and $\delta V_{\perp}$ (red)
vs $F_{D}$. 
(c) The corresponding fraction of six-fold coordinated skyrmions $P_{6}$  
vs $F_{D}$.
Dashed lines indicate the boundaries between the pinned state (phase I), 
the flow of interstitial skyrmions (phase II),
the disordered flow state (phase III), 
the segregated or phase-separated state
(phase PS), and the moving crystal (phase MC).
The letters a to d in panel (c) indicate the
values of $F_{D}$ at which the skyrmion images in Fig.~\ref{fig:2} were obtained.
}
\label{fig:1}
\end{figure}

\begin{figure}
\includegraphics[width=\columnwidth]{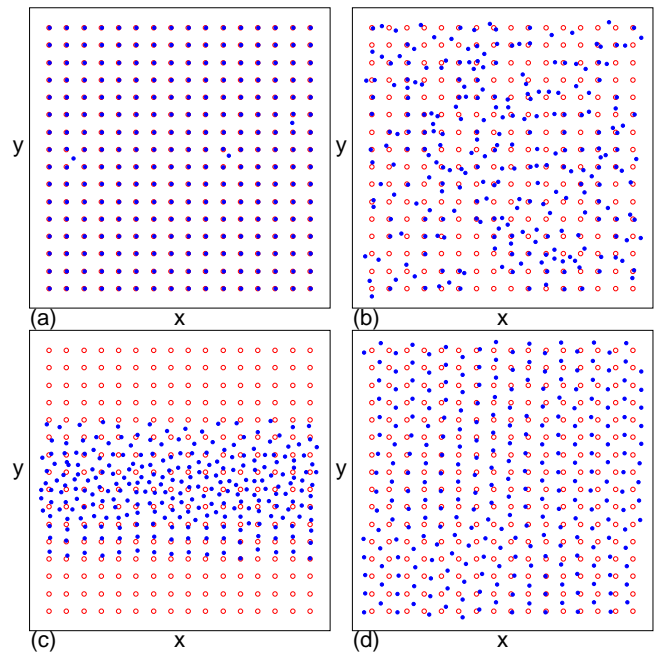}
\caption{Skyrmion (blue dots)
and pinning site (red circles) locations at one instant in time for the
system in Fig.~\ref{fig:1} with $f=1.0117$, $F_p=2.0$, $n_p=0.1975$,
and $\alpha_m/\alpha_d=9.96$.
(a) The pinned phase I at $F_{D} = 0.$
(b) The disordered flow phase III
at $F_{D} = 1.0$.
(c) The segregated or phase separated phase PS at $F_{D} = 3.0$.
(d) The moving crystal phase MC
at $F_{D} = 5.5$.   
}
\label{fig:2}
\end{figure}

In Fig.~\ref{fig:1}(a) we plot the velocity-force curve
$\langle V_{||}\rangle$ versus $F_{D}$ for a system  
with
$f=1.0117$,
a pinning density of $n_{p} = 0.1975$,
$F_{p} = 2.0$, and $\alpha_{m}/\alpha_{d} = 9.96$.
The $\langle V_{\perp}\rangle$ versus $F_D$ curve (not shown) has the
same shape but is larger in magnitude.
Figure~\ref{fig:1}(b) shows the mean square deviations in the instantaneous
velocities for both the parallel ($\delta V_{||}$) and perpendicular
($\delta V_{\perp}$) directions  versus $F_{D}$,
while Fig.~\ref{fig:1}(c) shows the corresponding fraction of six-fold
coordinated skyrmions $P_{6}$ vs $F_{D}$.
The vertical dashed lines
highlight the five different dynamic phases that arise.
Phase I is a pinned state with $\langle V_{||}\rangle = \langle V_{\perp}\rangle  = 0$.
As illustrated 
in Fig.~\ref{fig:2}(a)
at $F_{D} = 0,$
the skyrmions form a commensurate square lattice
with a small number of additional
skyrmions in the interstitial regions between pinning sites. 
Phase II consists of interstitial flow, in which only the interstitial
skyrmions depin and move around the
commensurate
skyrmions, which remain pinned.
The flow in phase II exhibits a series of directional locking effects which we explore
in detail in Section III A.
The value of $\langle V_{||}\rangle$ is small but finite, and there are
jumps in $\delta V_{||}$, $\delta V_{\perp}$, and $P_{6}$ at both ends of
the interval $0.15< F_{D} < 0.7$ where phase II occurs,
as shown in Fig.~\ref{fig:1}.
Within phase II the overall lattice structure is similar to that of phase I since
most of the skyrmions remain pinned.
Upward jumps in $\langle V_{||}\rangle$, $\delta V_{||}$, and $\delta V_{\perp}$ mark
the onset of the disordered flow phase III,
which appears in the interval
$0.7 \leq F_D \leq 2.01$, and consists of
a combination of pinned and flowing skyrmions
as
illustrated in Fig.~\ref{fig:2}(b) at $F_{D} = 1.0$.

Near $F_{D} = 2.0$ 
there is a transition to the clustered or segregated state,
shown in Fig.~\ref{fig:2}(c) at $F_{D} = 3.0$, where 
the skyrmions clump into a dense stripe.
We call this the segregated or phase separated phase PS.
There is an upward jump in $\delta V_{||}$ at the III-PS transition,
and $\delta V_{||}$
remains larger than
$\delta V_{\perp}$
throughout the PS phase
due to the alignment of the stripe with the driving direction.
This phase separated state
is very similar to that
observed
in 
continuum-based simulations of skyrmions moving over random pinning \cite{58},
and we find that the phase separation occurs only when the pinning is
sufficiently strong, in agreement with the continuum-based results.
The creation of the
segregated state in the continuum models was
attributed to the emission of
spin waves that produce
an effective attraction between the skyrmions \cite{58}.
In our system, no spin waves are present, and the segregation occurs
due to the velocity dependence of the skyrmion Hall angle in the presence of pinning.
If different regions of
the system are moving at different relative velocities,
each region will have a different skyrmion Hall angle, causing the
skyrmions in adjacent regions to gradually move toward one another.
In Fig.~\ref{fig:2}(c),
the bottom three rows of the stripe 
form a square lattice structure that is partially commensurate with
the pinning sites, so that in this region of the stripe,
the local filling factor is close to $f_{\rm loc}=1$.
The net skyrmion velocity is reduced close to commensuration,
so the value of $\theta_{sk}$ is smaller in this region.
The upper region of the stripe
is denser with $f_{\rm loc}>1$, and this incommensurate state has
a higher net skyrmion velocity and a correspondingly larger value
of $\theta_{sk}$.
As a result, there are velocity and $\theta_{sk}$ differentials
across the sample,
causing the larger $\theta_{sk}$ skyrmions
to collide with the smaller $\theta_{sk}$ skyrmions.
The skyrmions develop a nonuniform density profile that permits the
commensurate and incommensurate regions to slide past each other in
the driving direction, preserving a nonuniform velocity profile and
producing an increased velocity variation $\delta V_{||}$ in the PS
phase, as shown in Fig.~\ref{fig:1}(b).
At larger values of 
$F_{D}$,
$\theta_{sk}$ begins to saturate as shown 
in previous simulations \cite{49,50,51,52,56,57} and experiments \cite{53},
destroying 
the spatial differential in $\theta_{sk}$ and allowing the skyrmions
to form
a moving crystal state (MC),
as
illustrated in Fig.~\ref{fig:2}(d) at $F_{D} = 5.5$.
The PS-MC transition is accompanied by
drops in $\delta V_{||}$ and $\delta V_{\perp}$, which become
nearly isotropic,
along with an upward jump in $P_{6}$ to a value just below $P_6=1.0$.
The drops in $\delta V_{||}$ and $\delta V_{\perp}$ in the
MC phase occur since the skyrmions are all moving at the same velocity.

\subsection{Directional locking}

\begin{figure}
\includegraphics[width=\columnwidth]{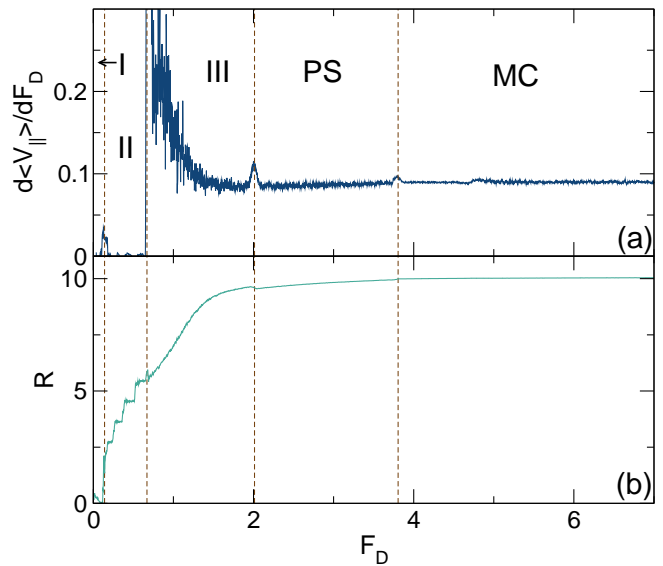}
\caption{ (a) $d\langle V_{||}\rangle/dF_{D}$ vs $F_{D}$
  for the system in Fig.~\ref{fig:1} with $f=1.0117$, $F_p=2.0$,
  $n_p=0.1975$, and $\alpha_m/\alpha_d=9.96$
  showing peaks at the transitions between different phases.
  I: pinned; II: interstitial flow; III: disordered flow; PS: phase separated;
  MC: moving crystal.
  (b) $R = \langle V_{\perp}\rangle/\langle V_{||}\rangle$
  vs $F_D$ for the same system.
}
\label{fig:3}
\end{figure}

In Fig.~\ref{fig:3}(a,b) we plot
the differential mobility
$d\langle V_{||}\rangle/dF_{D}$ and
the velocity ratio $R=\langle V_{\perp}\rangle/\langle V_{||}\rangle$
versus $F_D$
for the system
in Fig.~\ref{fig:1}.
Pronounced peaks appear in $d\langle V_{||}\rangle/dF_{D}$
across the I-II and II-III transitions, along with
smaller peaks
at the III-PS and PS-MC transitions.
The velocity ratio is zero
in phase I, and
increases in phases II and III
until it is close to the intrinsic value
$R_{\rm int}=\alpha_m/\alpha_d$ near
the onset 
of the PS state.
Within phase II, a series of steps appear in $R$
when the interstitial skyrmion flow
passes through a sequence of directionally locked states
as $F_{D}$ increases, similar to the
previously studied directional locking of skyrmion motion
on square arrays
in the single skyrmion limit \cite{50}. 
These locking effects occur due to the velocity dependence of the
skyrmion Hall angle
in the presence of
pinning, which causes the direction of skyrmion motion to rotate with
increasing $F_D$. 
Due to the underlying symmetry of the pinning potential
the rotation is not continuous; instead, the
motion becomes locked at integer
values of $R$ \cite{50}.
The steps in Fig.~\ref{fig:3}(b)
appear at $R = 2$, 3, 4, $5$ and $6$,
but all higher locking plateaus are
cut off when the system enters phase III.
Within phase III, $R$ increases smoothly,
and small dips in $R$ appear
at the III-PS and PS-MC transitions together
with small peaks in the 
differential
mobility
in Fig.~\ref{fig:3}(a).
     
\begin{figure}
\includegraphics[width=\columnwidth]{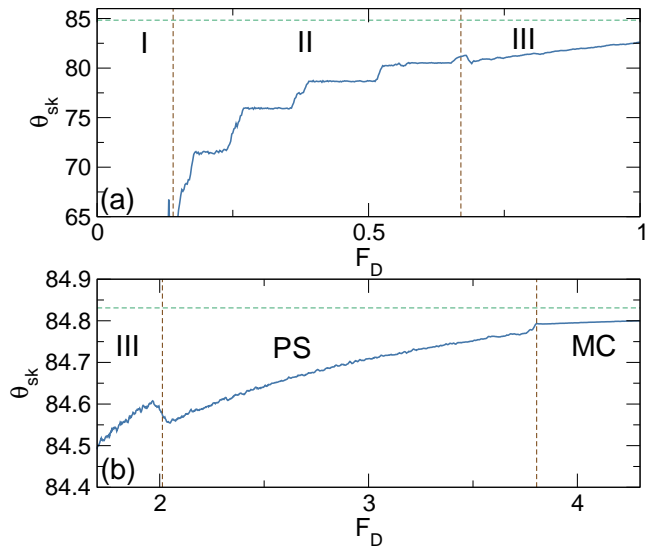}
\caption{ $\theta_{sk}$ vs $F_{D}$
  for the system in Fig.~\ref{fig:3}
  with $f=1.0117$, $F_p=2.0$, $n_p=0.1975$, and
  $\alpha_m/\alpha_d=9.96$,
  highlighting the directional locking
  steps in phase II (interstitial flow).
  (b) $\theta_{sk}$ vs $F_D$ for the same system
  in phases III (disordered flow), PS (phase separated), and MC (moving crystal),
  showing a jump down
  in $\theta_{sk}$ across the III-PS transition and a jump up
  across the PS-MC transition.
  The horizontal dashed lines in both panels indicate the
  disorder-free skyrmion Hall angle
  $\theta_{sk}^{\rm int}$.
}
\label{fig:4}
\end{figure}

In Fig.~\ref{fig:4}(a) we plot
$\theta_{sk} = \tan^{-1}(R)$ versus $F_{D}$,
showing more clearly that  
$\theta_{sk}$ is quantized in phase II.
The integer ratio $R = 3$ corresponds to the
step at $\theta_{sk} = 71.56^\circ$, and there are also steps for the $R=4$, $R=5$, and
$R=6$ ratios, with the latter
corresponding to the step at $\theta_{sk} = 80.53^\circ$. 
Much smaller steps appear at some rational fractional values of $R$
similar to
what was observed in the single skyrmion
limit \cite{50}. 
In Fig.~\ref{fig:4}(b) we show $\theta_{sk}$ versus $F_{D}$
across the III-PS and PS-MC transitions.
There is a drop in $\theta_{sk}$
at the III-PS transition caused when
the more slowly moving commensurate skyrmions
block the flow of the faster moving incommensurate
skyrmions, reducing the
net velocity in the perpendicular direction
and thereby reducing both $R$ and
$\theta_{sk}$.
When the PS structure destabilizes at the PS-MC transition,
the skyrmions begin to move at a uniform velocity in the MC phase,
eliminating the blocking of the flow of faster skyrmions by slower
skyrmions, and causing
$\theta_{sk}$ to increase.

\begin{figure}
\includegraphics[width=\columnwidth]{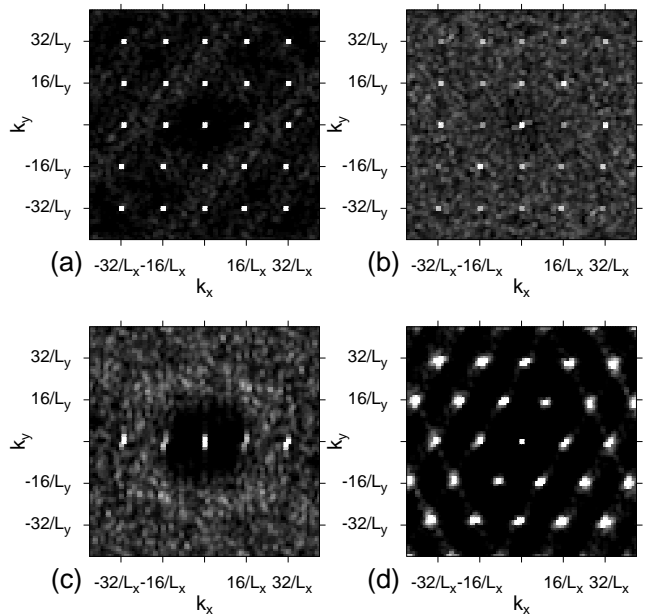}
  \caption{ The structure factor $S({\bf k})$
    for the system in Fig.~\ref{fig:1} with
    $f=1.0117$, $F_p=2.0$, $n_p=0.1975$, and $\alpha_m/\alpha_d=9.96$.
    (a) Pinned phase I.
    $S({\bf k})$ for the interstitial flow phase II
    (not shown) is
    similar in appearance.
    (b) Disordered flow phase III.
    (c) The PS phase.
    (d) The MC phase.
}
\label{fig:5}
\end{figure}

Some experimental techniques
cannot directly access the real-space position of the skyrmions but
instead measure
the structure factor  
$S({\bf k}) = N_{sk}^{-1}|\sum^{N_{sk}}_{i}\exp(-i{\bf k}\cdot{\bf r}_{i})|^2$.  
Figure~\ref{fig:5}(a) shows that $S({\bf k})$ in phase I
for the system in Fig.~\ref{fig:1}(a)  has square symmetry
since the skyrmions
are predominantly localized in the square pinning array.
In the interstitial flow phase II, $S({\bf k})$ also has strong square ordering
since most of the skyrmions remain pinned.
In the disordered flow phase III, where a combination of pinned and
flowing skyrmions appear,
$S({\bf k})$ exhibits a smearing due to the positional disorder, as shown
in Fig.~\ref{fig:5}(b);
however, there is still significant weight at the square lattice wavevectors
due to the pinned skyrmions.
In the PS phase illustrated in Fig.~\ref{fig:5}(c),
$S({\bf k})$ contains
a series of 
peaks along the $k_{y}$ axis,
indicative of smectic or stripe like ordering.
An additional
ring-like structure appears at larger values of ${\bf k}$
due to the partial disordering of the skyrmions within the stripe.
In the MC phase,
$S({\bf k})$
reveals strong triangular ordering,
as highlighted in Fig.~\ref{fig:5}(d).
These results show that the different
phases have distinct signatures in both real and reciprocal space,
and that transitions between the phases are associated with changes
in $\theta_{sk}$ and the transport properties.

\subsection{Varied pinning strength}

\begin{figure}
\includegraphics[width=\columnwidth]{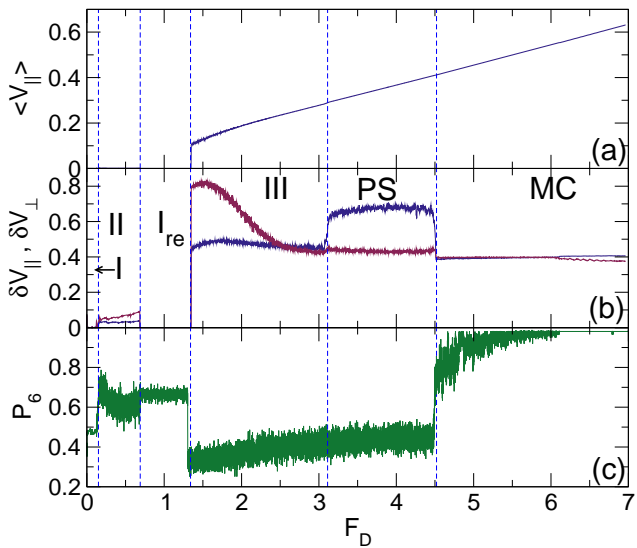}
\caption{  
  (a) $\langle V_{||}\rangle$ vs $F_{D}$
  for
  a system with $f=1.0117$, $n_p=0.1975$,
  $\alpha_m/\alpha_d=9.96$, and
  larger
  $F_p=3.0$.
  (b) The corresponding $\delta V_{||}$ (blue) and
  $\delta V_{\perp}$ (red) vs $F_{D}$. 
(c) The corresponding $P_{6}$ vs $F_{D}$.  
  Dashed lines indicate the boundaries between the phases.
  I: pinned, II: interstitial flow; I$_{re}$: reentrant pinned; III: disordered flow;
  PS: phase separated; MC: moving crystal.
}
\label{fig:6}
\end{figure}

We next consider the evolution of the phases for the system in Fig.~\ref{fig:1}
for varied pinning strength $F_{p}$.
In Fig.~\ref{fig:6}
we plot $\langle V_{||}\rangle$, $\delta V_{||}, \delta V_{\perp}$, and $P_{6}$
versus $F_{D}$ for
samples with $f=1.0117$, $n_p=0.1975$, and $\alpha_m/\alpha_d=9.96$ at
a larger
$F_{p} = 3.0$.
The III-PS and PS-MC transitions both shift to higher values of $F_D$ compared
to the $F_p=2.0$ case.
Although the overall behavior is similar for both values of $F_p$, in Fig.~\ref{fig:6}
we find
a
reentrant pinned phase falling between phases II and III over the range
$0.7 < F_{D} < 1.35$.
In the $F_D=0$ pinned phase I, 
the interstitial skyrmions
are pinned by the interactions with 
skyrmions located at the pinning sites.
The interstitial skyrmions depin and experience
a series of directional locking transitions in phase II,
but the stronger pinning sites are able to capture and repin the
interstitial skyrmions when the changing direction of motion causes
the skyrmions to pass directly over the pinning sites,
producing the reentrant pinned phase I$_{re}$.
For lower $F_{p}$, each
pinning site can capture only one skyrmion,
so the reentrant pinning does not occur;
however, for $F_{p} > 2.5$, a sufficiently large value of
$F_{D}$
can shift an already pinned skyrmion far enough to the edge of the
pinning site to permit a second skyrmion to enter the other side of the
pinning site and become trapped
in phase I$_{re}$.
At $F_{D} > 1.3$ for the $F_p=3.0$ system,
the drive is strong enough to depin
the doubly occupied pinning  
sites, and the system transitions to disordered phase III flow.  
In general,
we find that the same phases persist as $F_p$ is further increased, with the
width of the I$_{re}$ region increasing.
When $F_{p}$ becomes sufficiently large,
the interstitial skyrmions become trapped in doubly-occupied pinning sites
all the way down to $F_D=0$, causing phase II to disappear.
We also find that the jump in $P_{6}$ at the PS-MC transition
becomes less sharp for stronger $F_{p}$.

\begin{figure}
\includegraphics[width=\columnwidth]{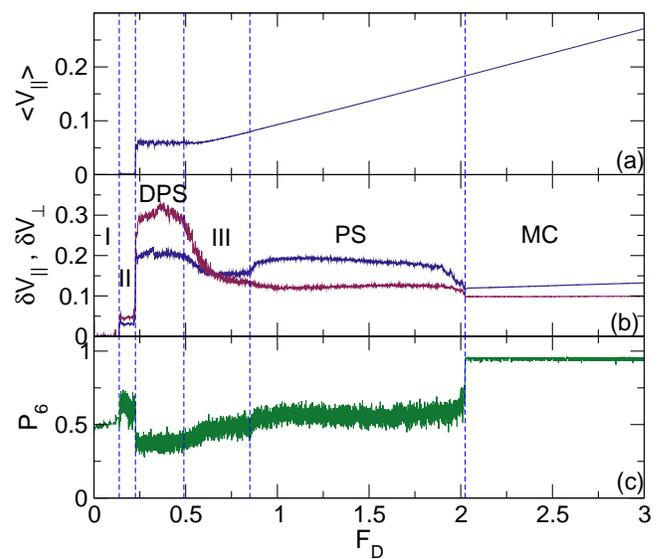}
\caption{ 
  (a) $\langle V_{||}\rangle$ vs $F_{D}$ for
  a system
  with $f=1.0117$, $n_p=0.1975$, and $\alpha_m/\alpha_d=9.96$
  at $F_p=0.75$ where a diagonal phase separated state, phase DPS, appears
  as illustrated in Fig.~\ref{fig:8}(a).
  (b) The corresponding $\delta V_{||}$ (blue) and
  $\delta V_{\perp}$ (red) vs $F_{D}$.
  (c) The corresponding $P_{6}$ vs $F_D$.
  Dashed lines indicate the phase boundaries.  I: pinned; II: interstitial flow; DPS:
  diagonal phase separated; III: disordered flow; PS: phase separated; MC:
  moving crystal.
}
\label{fig:7}
\end{figure}

\begin{figure}
\includegraphics[width=\columnwidth]{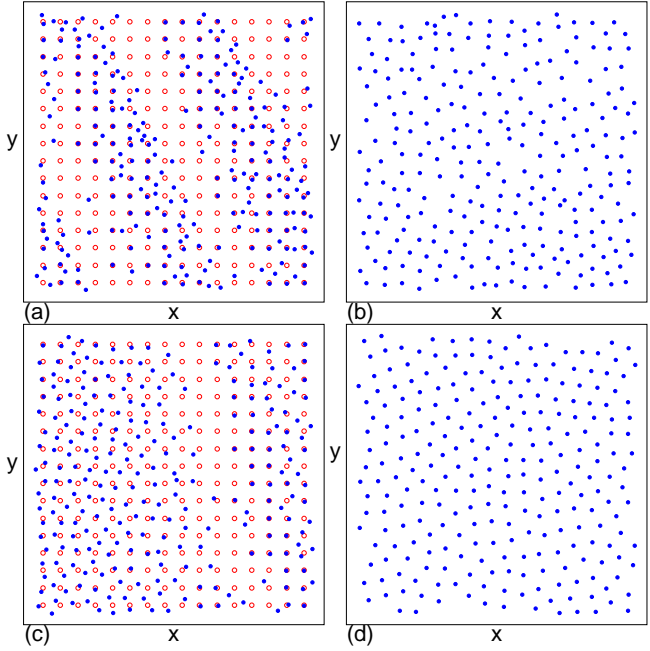}
\caption{
  (a,b) Skyrmion (blue dots) and pinning site (red circles) locations at one
  instant in time
  for the system in Fig.~\ref{fig:7}
  with $f=1.0117$, $n_p=0.1975$, $\alpha_m/\alpha_d=9.96$,
  and $F_p=0.75$.
  (a) The diagonal phase separated state DPS at $F_{D} = 0.25$,
  where multiple diagonal phase separated bands appear.
  (b) The
  phase separated PS state at $F_{D} = 1.5$,
  where the skyrmion density is more uniform.
  The pinning sites are not shown for clarity.
  (c,d) Skyrmion and pinning site locations for a system with
  $f=1.0117$, $n_p=0.1975$, $\alpha_m/\alpha_d=9.96$, and 
  $F_p=0.2$.
  (c) The DPS phase
  at $F_{D} = 0.075$, where only one band of skyrmions forms.
  (d) Phase III flow
  at
  $F_{D} = 0.55$.
  The pinning sites are not shown for clarity.
}
\label{fig:8}
\end{figure}

In Fig.~\ref{fig:7}
we
plot $\langle V_{||}\rangle$, $\delta V_{||}, \delta V_{\perp}$, and $P_{6}$
versus $F_{D}$ for the same system in Fig.~\ref{fig:6} but at  
a weaker $F_{p} = 0.75$.
We again observe
phases
I, II, III, PS, and MC;
however,
a diagonal phase separated state
called phase DPS
now appears between phases II and III.
In
the DPS phase,
$\langle V_{||}\rangle$ gradually decreases
with increasing $F_{D}$ as 
shown in Fig.~\ref{fig:7}(a). 
We illustrate the DPS flow at $F_D=0.25$ in Fig.~\ref{fig:8}(a),
where the skyrmions form a series of dense bands
at an angle to the driving direction.
One distinction between the DPS and PS phases
is that in the DPS phase
there is a combination of pinned skyrmions and moving skyrmions,
whereas in the PS phase,
all the skyrmions are moving.   
At higher driving the system
undergoes
uniform disordered phase III flow
before transitioning into the PS phase
shown in 
Fig.~\ref{fig:8}(b) for
$F_{D} = 1.5$.
In general, as $F_{p}$ decreases,
the width of the stripe in the PS phase increases.
Near $F_{D} = 2.0$, the system transitions into the MC phase.

\begin{figure}
\includegraphics[width=\columnwidth]{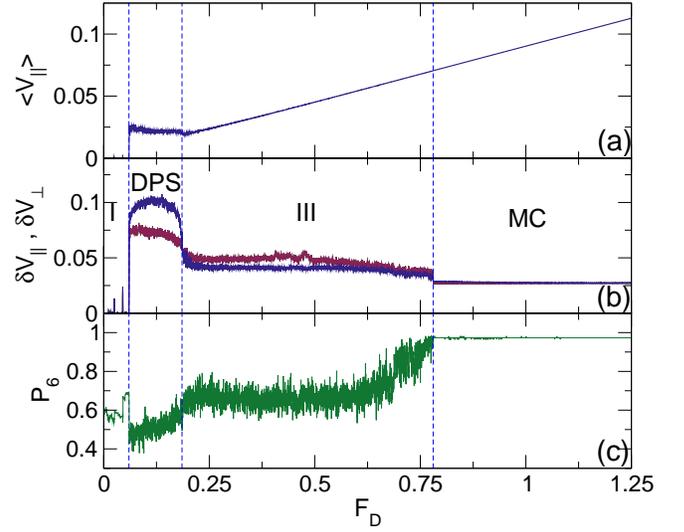}
\caption{  
  (a) $\langle V_{||}\rangle$ vs $F_{D}$
  for
  a system
  with $f=1.0117$, $n_p=0.1975$, and $\alpha_m/\alpha_d=9.96$
  at  $F_{p} = 0.2$.
  Here phase II  is lost and the system depins directly from
  phase I to phase DPS.
  (b) The corresponding $\delta V_{||}$ (blue)
  and $\delta V_{\perp}$ (red)  vs $F_{D}$.
  (c) The corresponding $P_{6}$
  vs $F_{D}$.
  Dashed lines indicate the phase boundaries.  I: pinned; DPS: diagonal phase
  separated; III: disordered flow; MC: moving crystal.
}
\label{fig:9}
\end{figure}

\begin{figure}
\includegraphics[width=\columnwidth]{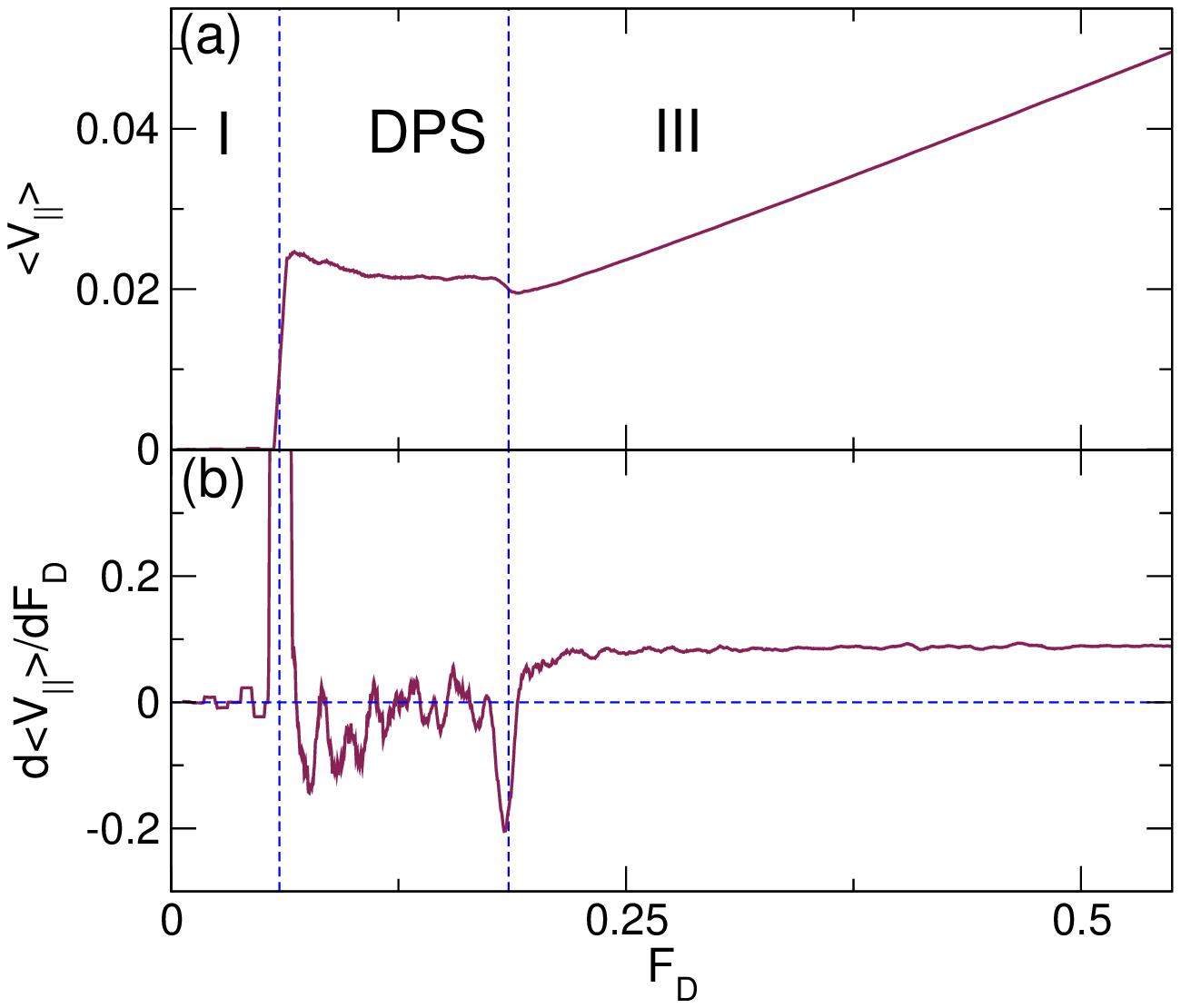}
\caption{ (a) A blowup of $\langle V_{||}\rangle$ vs $F_{D}$ for the
  system in Fig.~\ref{fig:9} with $f=1.0117$,
  $n_p=0.1975$, $\alpha_m/\alpha_d=9.96$, and $F_p=0.2$
  in the region containing the I-DPS and DPS-III transitions.
  A dip in $\langle V_{||}\rangle$
  appears at the DPS-III transition,
  producing
  negative differential conductivity,
  $d\langle V_{||}\rangle/dF_{D} < 0$.
  (b) $d\langle V_{||}\rangle/dF_{D}$ vs $F_D$ for the same system.
  For clarity, this
  data
  has been subjected to a running average in order
  to reduce the noise.
}
\label{fig:10}
\end{figure}

In Fig.~\ref{fig:9} we plot  $\langle V_{||}\rangle$,
$\delta V_{||}, \delta V_{\perp}$, and $P_{6}$
versus $F_{D}$
for the system in Fig.~\ref{fig:6}
at $F_{p} = 0.2$,
where phase II is lost and the system transitions
directly from phase I to
phase DPS.
At lower values of $F_{p}$ such as this, the repulsion
between the interstitial skyrmions and
the skyrmions at the pinning sites is strong enough
that when the interstitial skyrmions depin, their motion causes
the skyrmions in the neighboring pinning sites to depin as well,
destroying the interstitial flow phase II.
The PS phase is also lost
and phase III is followed immediately
by the MC phase. 
As $F_p$ decreases,
the spatial width  of the
bands in the DPS phase
increases, as illustrated
in Fig.~\ref{fig:8}(c) for the system in
Fig.~\ref{fig:9} at $F_{D} = 0.075$.
The disordered flow phase III in the same system 
is much more uniform 
than the
phase III flow found at higher values of $F_{p}$,
as shown in Fig.~\ref{fig:8}(d) at $F_D=0.55$.
The DPS-III transition is relatively sharp and appears
as a jump down
in $\delta V_{||}$ and $\delta V_{\perp}$
along with a jump up in $P_{6}$.
There is also a  dip in 
$\langle V_{||}\rangle$ indicating a drop
in the average velocity of the skyrmions in the direction of drive at the DPS-III
transition,
as shown more clearly
in the zoomed-in plot of Fig.~\ref{fig:10}(a).
This dip
creates a regime of negative differential conductivity
where $d\langle V_{||}\rangle/dF_{D} < 0$,
illustrated in Fig.~\ref{fig:10}(b).

\begin{figure}
\includegraphics[width=\columnwidth]{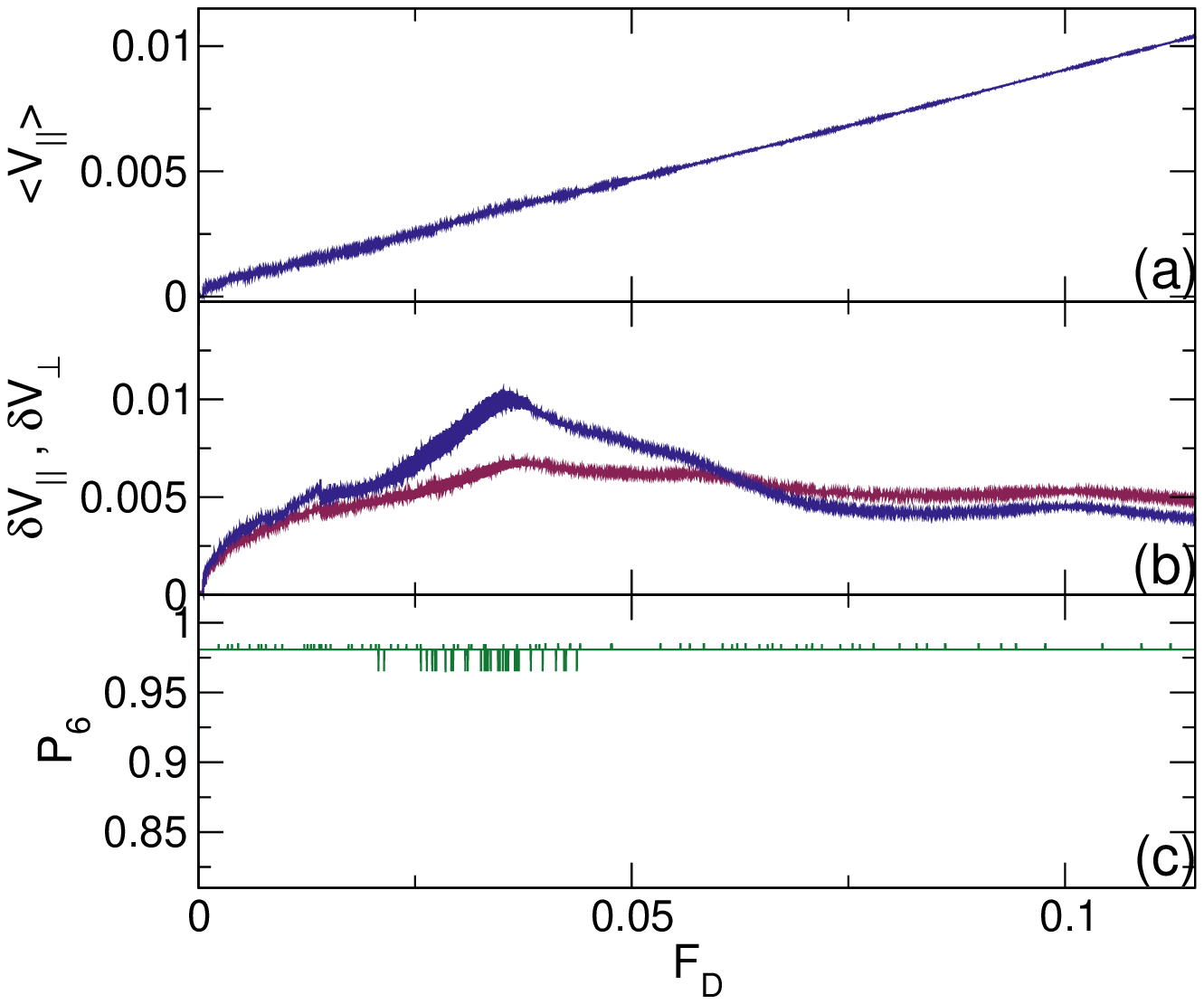}
\caption{  
  (a) $\langle V_{||}\rangle$ vs $F_{D}$ for
  a system
  with $f=1.0117$, $n_p=0.1975$, and $\alpha_m/\alpha_d=9.96$
  at $F_{p} = 0.015$, 
  where the
  skyrmions depin elastically from a floating triangular solid to a MC state.  
(b) The corresponding $\delta V_{||}$ (blue)
and $\delta V_{\perp}$ (red) vs $F_{D}$.
(c) The corresponding $P_{6}$ vs $F_{D}$.
}
\label{fig:11}
\end{figure}

When $F_{p} < 0.02$,
in the absence of driving there is a transition from a
commensurate solid where the skyrmions
are predominantly located at the pinning sites
to a floating solid where the skyrmions form a triangular
lattice that is only weakly coupled to the pinning lattice.
This transition is accompanied by a 
drop in the depinning threshold force $F_{c}$.
In Fig.~\ref{fig:11}(a,b,c)
we show $\langle V_{||}\rangle$,
$\delta V_{||}$, $\delta V_{\perp}$, and $P_{6}$
versus $F_D$
for the system in Fig.~\ref{fig:6} 
at $F_{p} = 0.015$.
Here
$P_{6}$ is close to $1.0$ for all values of $F_{D}$,
so the
skyrmions depin directly 
from a weakly pinned floating solid into a moving crystal. 
The depinning transition is elastic
so all of the 
skyrmions keep their same neighbors.
Additionally, both $\delta V_{||}$ and $\delta V_{\perp}$ are small and
vary only weakly with
$F_{D}$
due to the
slow rotation of the skyrmion lattice with
increasing $F_{D}$
that occurs
as
the triangular
lattice
remains
oriented with the direction of motion.

\begin{figure}
\includegraphics[width=\columnwidth]{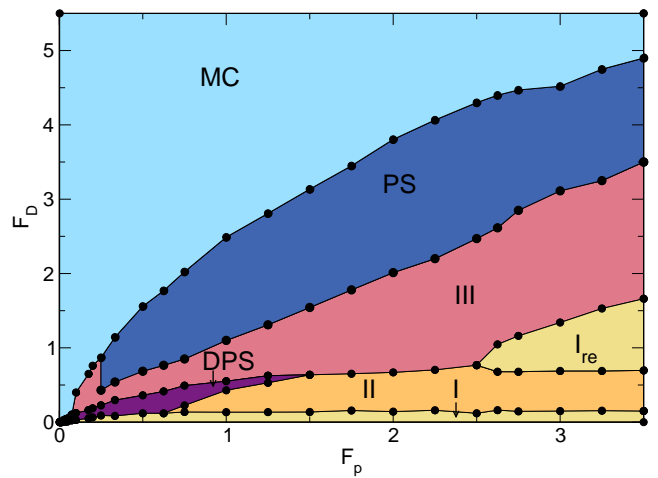}
\caption{ 
  Dynamic phase diagram as a function of $F_D$ vs $F_p$
  for the system in Fig.~\ref{fig:1}
  with $f=1.0117$, $n_p=0.1975$, and $\alpha_m/\alpha_d=9.96$, showing the
  pinned phase I, the interstitial flow phase II,
  the reentrant pinned phase I$_{re}$,
  the disordered phase III,
  the phase separated state
  PS,
  the diagonal phase separated state DPS, and the moving crystal state MC.  
}
\label{fig:12}
\end{figure}

By conducting a series of simulations for varied
$F_{p}$ and $F_{D}$, and examining the changes in the
transport signatures and structural properties, 
we can construct a dynamical phase diagram
as a function of $F_D$ versus $F_p$, 
as shown in Fig.~\ref{fig:12} for a system with
$f=1.0117$, $n_p=0.1975$, and $\alpha_m/\alpha_d=9.95$.
The PS phase only occurs for $F_{p} > 0.25$. This is consistent with the results
found in continuum simulations,
where a moving segregated state only appears when the disorder 
is sufficiently strong \cite{58}.
The PS-MC transition shifts to higher values of $F_D$ 
with increasing $F_{p}$,
and the extent of phase III also grows with increasing $F_{p}$.
For $F_{p} > 2.5$,
a reentrant pinned phase appears and grows in extent with increasing $F_p$.
Phase II only occurs when $F_{p} > 0.65$, and
the extent of phase II saturates as $F_p$ increases
since the transition out of phase II on the high drive side
is controlled by the skyrmion-skyrmion interaction
potential, which remains constant.
Phase DPS
occurs in the window
$0.075 < F_{p} < 1.5$.
The results in Fig.~\ref{fig:12}
indicate
that there is a wide range of pinning strength
over which some
form of dynamical phase separation of the driven skyrmions can be realized.
Similar phase segregation can occur for
strong random pinning, as will be described elsewhere.

\begin{figure}
\includegraphics[width=\columnwidth]{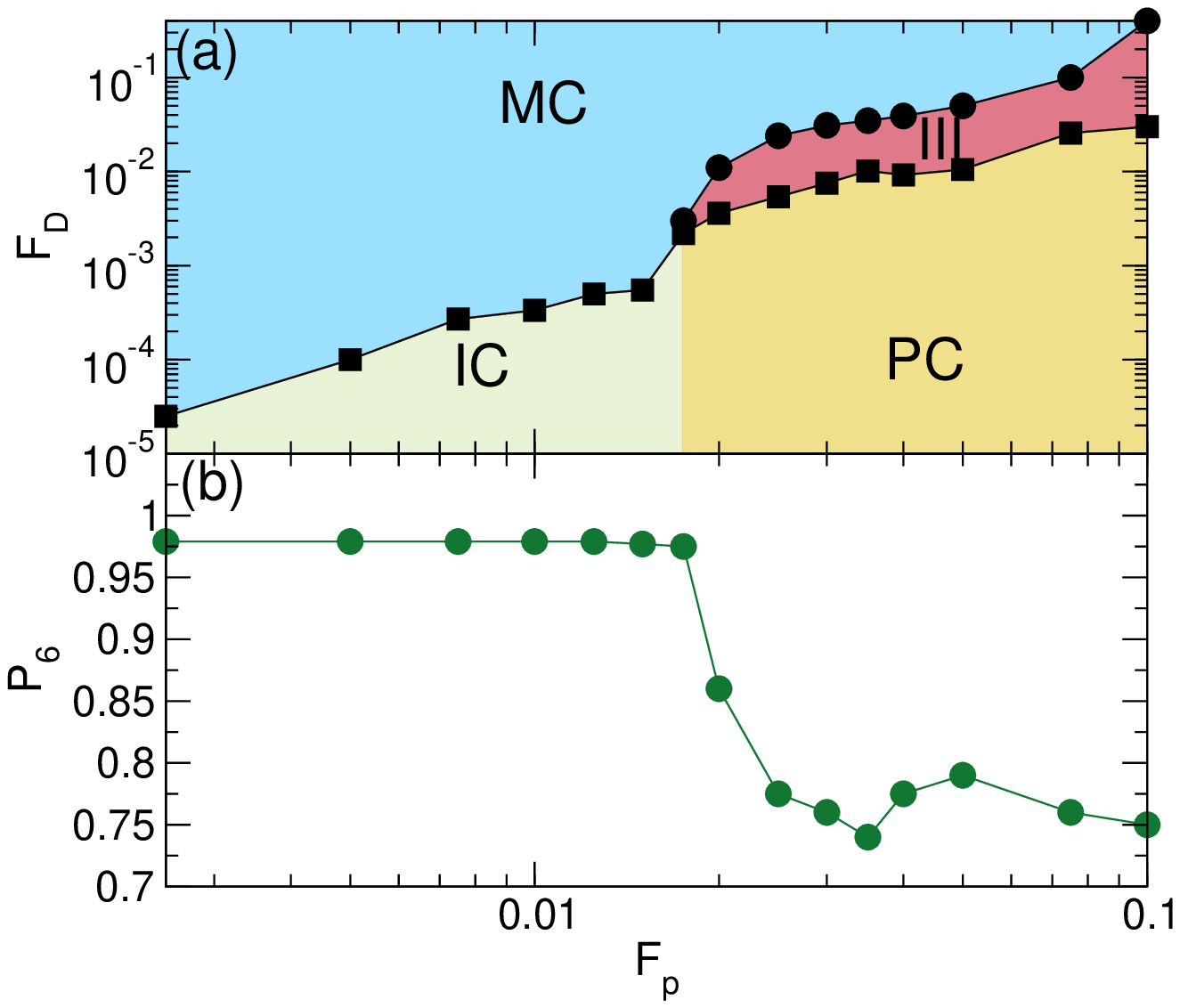}
\caption{ 
  (a) A blow up of the weak pinning region
  in the dynamic phase diagram from Fig.~\ref{fig:12} as a function of $F_D$ vs $F_p$
  showing the transition in the pinned phase I
  from a pinned commensurate solid (PC) to
  an incommensurate floating solid (IC) near $F_{p} = 0.02$. This
  transition also produces a jump down in $F_{c}$.
  (b) $P_{6}$ vs $F_{p}$ at $F_{D} = 0$ in the same system
  showing  a drop in $P_{6}$ across the IC-PC line. 
}
\label{fig:13}
\end{figure}

In Fig.~\ref{fig:13}(a) we highlight
a portion of the dynamic phase diagram from Fig.~\ref{fig:12} in
the weak substrate regime of  $F_{p} < 0.1$. 
For $F_{p} < 0.02$, the skyrmions form an
incommensurate floating triangular crystal,
called the IC phase.  The pinned phase I described above is marked as a
pinned commensurate solid, phase PC, in Fig.~\ref{fig:13}(a).
The PC-IC transition is associated
with a drop in $F_{c}$ since the IC depins elastically into the MC phase, while
the PC depins plastically and passes through disordered phase III flow before
reaching the MC state.
In Fig.~\ref{fig:13}(b) we plot $P_{6}$ vs 
$F_{p}$ at $F_{D} = 0$,
showing that $P_6$ drops across the IC-PC transition
from
$P_6 \approx 1.0$ in the IC state to
$P_6 \approx 0.75$ in the PC state.
The dynamic phase diagram
shown in Fig.~\ref{fig:13}(a)
is similar
to
that observed for weak random pinning,
with the pinned crystal depinning elastically into a moving crystal,
where the pinned commensurate crystal in the periodic pinning array is replaced
by a disordered pinned glassy state in the random pinning array
\cite{49}.
We note that
Ref.~\cite{49} focused on random pinning in the limit below the pinning strength
at which dynamical segregation occurs.
The PC-IC transition
and the associated drop in $F_{c}$
are similar to
what is observed at an
Aubry transition \cite{59} of the type found
in colloidal systems 
with periodic substrates as the coupling
of the colloids to the substrate is decreased \cite{60}. 

\section{Changing the Strength of the Magnus Force}

\begin{figure}
\includegraphics[width=\columnwidth]{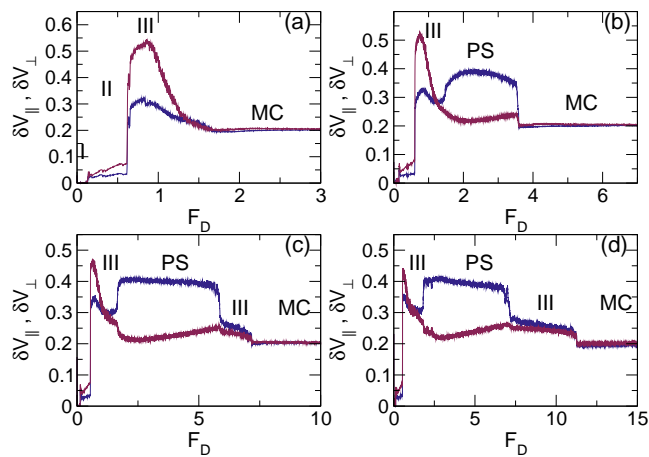}
\caption{ 
  $\delta V_{||}$ (blue) and $\delta V_{\perp}$ (red)
  vs $F_{D}$ for
  a system with $f=1.0117$, $n_p=0.1975$, and 
 $F_{p} = 1.5$. 
  (a) $\alpha_{m}/\alpha_{d} = 6.591$, where there is no PS phase.
  (b) $\alpha_{m}/\alpha_{d}= 13.3$. 
  (c) $\alpha_{m}/\alpha_{d} = 26.7$, where there is a
  reentrant phase III above the PS phase.
  (d) $\alpha_{m}/\alpha_{d} = 39.98$.
  I: pinned; II: interstitial flow; III: disordered flow; PS: phase separated;
  MC: moving crystal.
}
\label{fig:14}
\end{figure}

We next consider how the dynamical phases evolve when the pinning strength
is held fixed but the ratio of the Magnus term to the dissipative term is varied.
We use $f=1.0117$ and $n_p=0.1975$, and
set 
$F_{p} = 1.5$, since as indicated in Fig.~\ref{fig:12}, phases
I, II, III, PS, and MC all appear for this pinning strength when
$\alpha_{m}/\alpha_{d} = 9.96$.
In Fig.~\ref{fig:14} we plot
$\delta V_{||}$ and $\delta V_{\perp}$ versus $F_{D}$ since these measures
best highlight the 
transition between the different phases.
At $\alpha_{m}/\alpha_{d} = 6.591$ in Fig.~\ref{fig:14}(a),
phases I, II, III and MC are present while the PS phase is absent,
and in general for this value of $F_{p}$,
the PS phase only occurs for $\alpha_{m}/\alpha_{d} > 7.0$. 
In Fig.~\ref{fig:14}(b)
at $\alpha_{m}/\alpha_{d} = 13.3$,
all five phases appear and the PS-MC transition shifts up
to $F_{D} = 3.65$.
Figure~\ref{fig:14}(c) shows
$\alpha_{m}/\alpha_{d} = 26.7$,
where the PS phase is larger in extent and a reentrant
phase III appears
between the PS and the MC phases.
At 
$\alpha_{m}/\alpha_{d} = 39.98$ in
Fig.~\ref{fig:14}(d),
both the PS phase and the reentrant phase III have increased in extent.
In each case, in the PS phase $\delta V_{||}$ is always larger than 
$\delta V_{\perp}$, while in the MC phase,
$\delta V_{||} \approx \delta V_{\perp}$.

\begin{figure}
\includegraphics[width=\columnwidth]{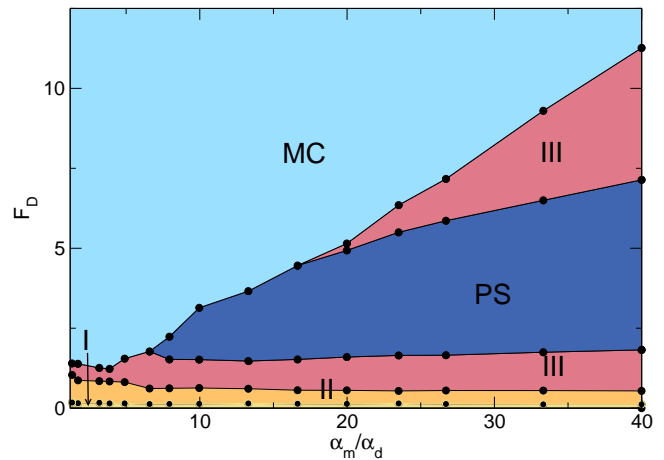}
\caption{ 
  Dynamic phase diagram as a function of
  $F_{D}$ vs $\alpha_{m}/\alpha_{d}$ for  the system in Fig.~\ref{fig:12}
  with $f=1.0117$ and $n_p=0.1975$ at $F_{p} = 1.5$. 
  Here the PS phase only occurs for $\alpha_{m}/\alpha_{d} > 7.0$.
  There is also a reentrant phase III that appears when
  $\alpha_{m}/\alpha_{d} > 15$.
  I: pinned; II: interstitial flow; III: disordered flow; PS: phase separated;
  MC: moving crystal.
}
\label{fig:15}
\end{figure}

In Fig.~\ref{fig:15} we show a dynamic phase diagram
for the system in Fig.~\ref{fig:14} at $F_p=1.5$
as a function of $F_D$ versus $\alpha_m/\alpha_d$, where we
highlight phases I, I, III, PS, and MC.
The value of $F_D$ at which the MC phase
appears increases linearly with increasing Magnus force
for $\alpha_{m}/\alpha_{d} > 4.0$, 
while for $\alpha_{m}/\alpha_{d} \leq 4.0$ it 
remains roughly constant. 
The PS phase occurs when $\alpha_{m}/\alpha_{d} > 7.0$,
and it grows in extent with increasing $\alpha_{m}/\alpha_{d}$, indicating
that the PS phase is produced by the
dynamics associated with the Magnus force. 
The I-II transition occurs at a nearly constant value of $F_D$
since the depinning of the interstitial skyrmions is determined by the magnitude of the
skyrmion-skyrmion interaction force,
which is independent of $\alpha_{m}/\alpha_{d}$.
In contrast,
the II-III transition shifts to lower $F_D$ as
$\alpha_{m}/\alpha_{d}$ increases from zero,
since an increase in the Magnus force gives the skyrmions a stronger
tendency to move in the direction perpendicular to the drive,
increasing the probability that a moving interstitial skyrmion will depin a
pinned skyrmion and produce the onset of disordered phase III flow.
For $\alpha_{m}/\alpha_{d} \leq 1.0$,
a different
set of dynamical phases arise that are
similar to those
found in previous simulations of
vortices interacting with periodic pinning \cite{15,16,30},
since the behavior in this limit
is associated with strong damping.
A detailed study of this regime will appear elsewhere.

\section{Varied Filling Factor}

\begin{figure}
\includegraphics[width=\columnwidth]{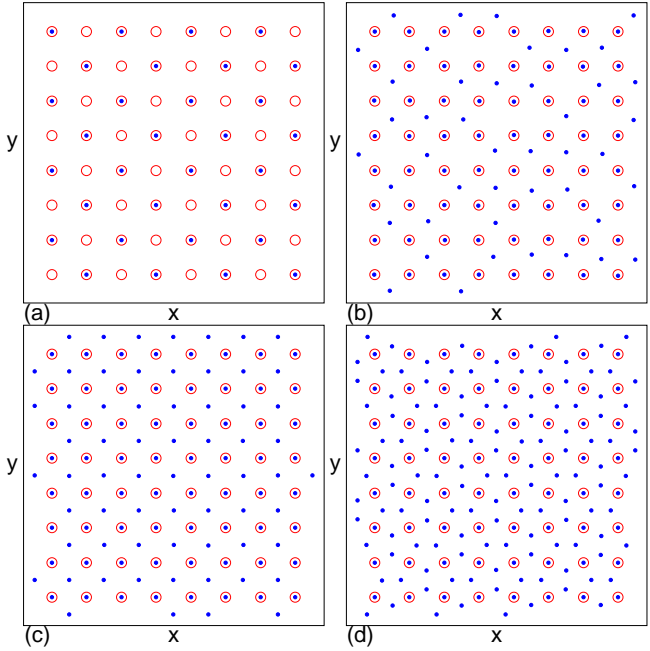}
\caption{Skyrmion (blue dots) and pinning site (red circles)
  locations for the system in Fig.~\ref{fig:1}
  with $n_p=0.1975$ and $\alpha_m/\alpha_d=9.96$
  at $F_{p} = 2.0$ and $F_{D} = 0$.
  (a) $f = 0.5$, where an ordered checkerboard state appears.
  (b)  $f = 1.65$.
  (c) $f = 2.0$, where
  there is another ordered state.
  (d)  $f=3.0$, where a partially ordered dimer state occurs.
}
\label{fig:16}
\end{figure}

We next consider the evolution of the different phases when the ratio of the
number of skyrmions to the number of pinning sites is varied.
We hold the pinning density fixed at
$n_{p} = 0.1975$, and
set $F_{p} = 2.0$ and $\alpha_{m}/\alpha_{d} = 9.96$.
In Fig.~\ref{fig:16}
we show the $F_D=0$ skyrmion and pinning site
locations at different filling factors.
At $f = 0.5$ in Fig.~\ref{fig:16}(a), 
the skyrmions form a commensurate checkerboard pattern in which
every other pinning site is occupied.
A partially disordered state appears for
$f = 1.65$ in Fig.~\ref{fig:16}(b).
At $f = 2.0$ in Fig.~\ref{fig:16}(c),
we find an ordered commensurate square lattice
with one skyrmion per pinning site and one skyrmion in each interstitial
plaquette.
Figure~\ref{fig:16}(d) shows that at
$f = 3.0$,
there is a partially ordered dimer state.
In general, the $F_D=0$ ordering of the skyrmions as function of
filling factor on a square pinning lattice is same as that
observed
for vortices in type-II superconductors with square pinning arrays \cite{9,11,25,26,27}.
Although the dynamics of the skyrmions is different 
from that of the vortices, the Magnus force has no effect on the static pinned
configurations.
The skyrmion arrangements
in Fig.~\ref{fig:16} were
obtained through simulated annealing; however, in
an experimental
system,
pinned configurations could be prepared using other means such as 
a rapid quench.
In that case, the Magnus dynamics would be relevant to the quench dynamics and
could alter 
the pinned configurations compared to those of quenched superconducting vortices.

\begin{figure}
\includegraphics[width=\columnwidth]{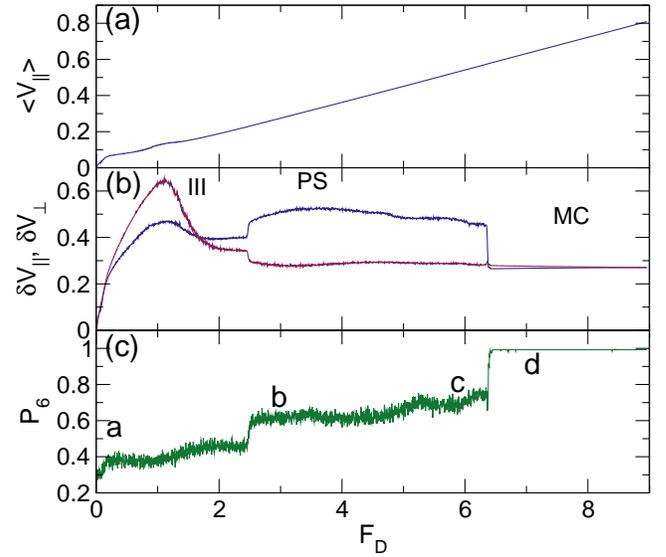}
\caption{(a) $\langle V_{||}\rangle$ vs $F_{D}$ for the system in Fig.~\ref{fig:16}
    with $n_p=0.1975$, $\alpha_m/\alpha_d=9.96$, $F_p=2.0$ and $f=2.5$.
  (b) The corresponding $\delta V_{||}$ (blue) and $\delta V_{\perp}$ (red)
  vs $F_{D}$.  III: disordered flow; PS: phase separated; MC: moving crystal.
  (c)
  The corresponding $P_{6}$ vs $F_{D}$.
  The letters a to d
  indicate the values of $F_D$ at which the images
  in Fig.~\ref{fig:18} were obtained.
}
\label{fig:17}
\end{figure}

\begin{figure}
\includegraphics[width=\columnwidth]{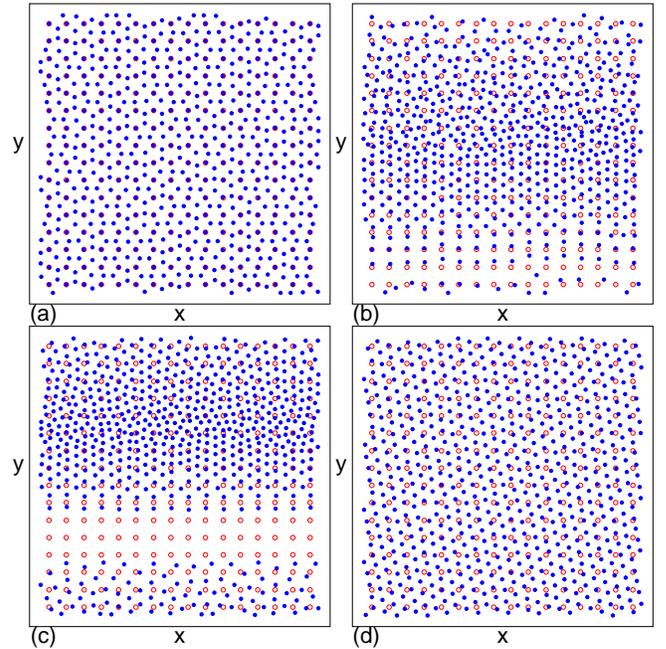}
\caption{Skyrmion (blue dots) and pinning site (red circles) locations at
  one instant in time
  for the system in Fig.~\ref{fig:17}
  with $n_p=0.1975$, $\alpha_m/\alpha_d=9.96$, and $F_p=2.0$
  at
  $f = 2.5$. 
(a) A uniform distribution of skyrmions and pinning sites at $F_{D} = 0.0$. (b)
The PS phase at $F_{D} = 2.75$. (c) The latter part of the PS phase at $F_{D} = 6.0$. (d) 
The MC phase at $F_{D} = 7.25$.
Note that due to the periodic boundary conditions, the upper portion of the stripe
region in panel (c) appears at the bottom of the figure.
}
\label{fig:18}
\end{figure}

In general, we find that the PS phase appears
for filling factors of 
$f > 0.6$,
and that the extent of the PS phase increases as
$f$ increases.
In Fig.~\ref{fig:17}
we plot $\langle V_{||}\rangle$,
$\delta V_{||}$, $\delta V_{\perp}$, and $P_{6}$ versus $F_{D}$ for
a sample with
$f = 2.5$, showing the extended range of the PS phase.
At the higher filling factors,
some 
additional smaller changes
occur in the PS phase,
such
as the increase in $P_{6}$ near $F_{D} = 5.0$ in Fig.~\ref{fig:17}(c),
where also correlates with a smaller decrease in $\delta V_{||}$
in Fig.~\ref{fig:17}(b).
There is also a small jump in $\langle V_{||}\rangle$ which
correlates with a peak in
$\delta V_{||}$ and $\delta V_{\perp}$ near $F_{D} = 1.0$.
Figure~\ref{fig:18}(a) shows the pinned configuration at $F_{D} = 0$,     
where we find
a partially ordered incommensurate state
instead of
the square lattice
that appears near $f = 1.0$.
In Fig.~\ref{fig:18}(b) we illustrate
the skyrmion configurations at $F_{D} = 2.75$ in the
PS phase.
A skyrmion density gradient forms,
with 
a square skyrmion lattice at a
local filling factor $f_{\rm loc} \approx 1$
in the lower portion of the stripe,
a denser square skyrmion lattice 
rotated by $45^\circ$ with respect to the pinning lattice 
at a filling factor $f_{\rm loc} \approx 2$ between the lower and middle portions of the
stripe,
a disordered high density skyrmion arrangement in the center of the stripe,
and a lower density disordered skyrmion arrangement at the top of the stripe.
The density profile in Fig.~\ref{fig:18}(b) suggests that there should
be steps in the skyrmion density corresponding to
integer local filling factors.
Similar steps in the particle density were
previously  proposed to occur in a density gradient of
superconducting vortices 
on periodic pinning arrays \cite{63,64}, and such steps
also have similarities
to the so-called wedding cake density profiles
for cold atoms on periodic optical lattices \cite{65}.
The distinction between these other systems and the skyrmion system is that, for
the skyrmions, the effect is entirely dynamical in nature and occurs only in the
moving state.
If the drive were suddenly removed, the PS structure would rapidly relax back into
a uniform pinned crystal state.
Figure~\ref{fig:18}(c) shows the skyrmion structure
at a higher drive of $F_D=6.0$ in 
the PS phase.
The step
feature in the lower
portion of the stripe is diminished
while the upper
portion of the stripe is now more ordered,
which is why $P_{6}$ is slightly higher at
this value of $F_{D}$.
The ordering in the upper half of the stripe (which is most apparent on the bottom of
the figure due to the periodic boundary conditions)
also shows characteristic arcs similar to those found in conformal crystals,
such as two-dimensional crystals of repulsive particles subjected to
a density gradient
\cite{66,67,68}. When the drive is high enough, 
the velocity differential between the skyrmions diminishes until
the clustering instability is lost and the system forms
a triangular lattice, as illustrated in
Fig.~\ref{fig:18}(d) at $F_{D} = 7.25$.

\begin{figure}
\includegraphics[width=\columnwidth]{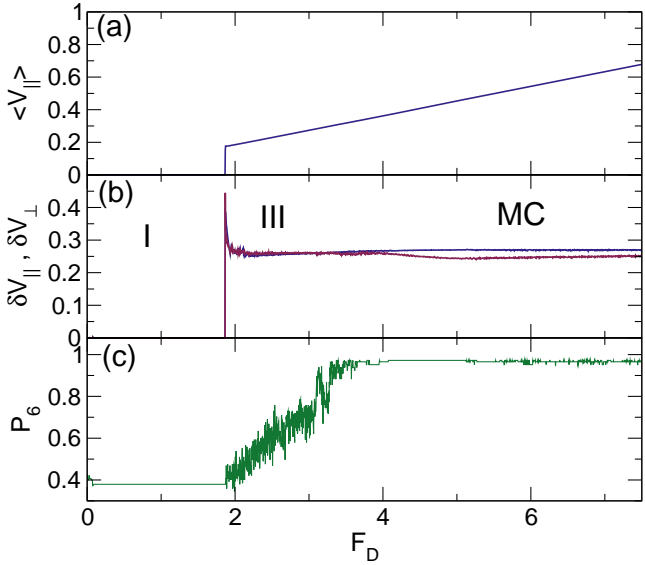}
\caption{
  (a) $\langle V_{||}\rangle$ vs $F_{D}$ for the system in Fig.~\ref{fig:16}
  with $n_p=0.1975$, $\alpha_m/\alpha_d=9.96$, and $F_p=2.0$
  at a filling
  factor of
    $f = 0.56$, where only phases I (pinned), III (disordered flow), and MC
  (moving crystal) appear.
  (b) The corresponding
  $\delta V_{||}$ (blue) and $\delta V_{\perp}$ (red) vs $F_{D}$.
  (c) The corresponding $P_{6}$ vs $F_{D}$.
}
\label{fig:19}
\end{figure}

\begin{figure}
\includegraphics[width=\columnwidth]{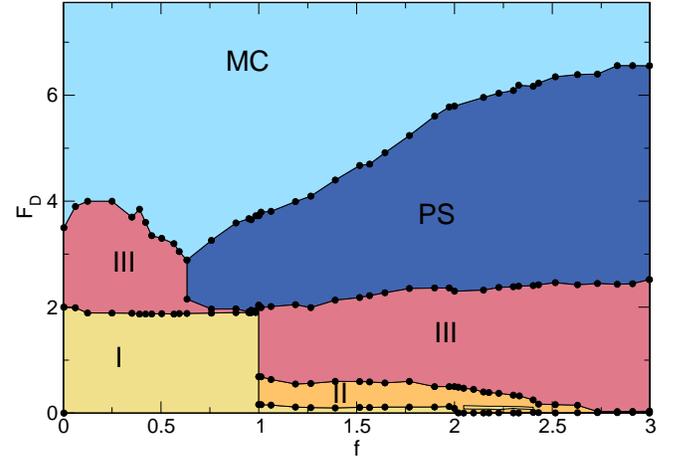}
\caption{Dynamic phase diagram as a function of $F_{D}$ vs
  $f$ for
  the system in Fig.~\ref{fig:16}
  with
  $n_p=0.1975$, $\alpha_m/\alpha_d=9.96$, and
  $F_{p} = 2.0$.
  For clarity, the reentrant phase I region falling between  $2.04 \leq f \leq 2.4$ is
  marked only by lines and not with data points.
  I: pinned; II: interstitial flow; III: disordered flow; PS: phase separated;
  MC: moving crystal.
}
\label{fig:20}
\end{figure}

In Fig.~\ref{fig:19} we plot
$\langle V_{||}\rangle$,
$\delta V_{||}$, $\delta V_{\perp}$, and $P_{6}$ versus $F_{D}$ for the
system in Fig.~\ref{fig:16} at
$f=0.56$, where
only three phases appear.  Phases I, III, and MC are present while phases II and PS
are lost.
At low skyrmion densities such as this,
the skyrmion-skyrmion interactions
are weak,
so that even if there are regions where skyrmions are moving 
faster or slower than average, skyrmions from other regions can easily move
around them, preventing the density gradient required to produce the PS phase
from occurring.
Figure~\ref{fig:20} shows the dynamic phase diagram
as a function of $F_{D}$ versus filling factor
$f$ for the system in Fig.~\ref{fig:16}.
A large increase in the pinned phase I occurs when $f<1$
due to the disappearance of skyrmions from the interstitial locations,
so that all of the skyrmions in the system are directly pinned by pinning
sites.
There are small peaks in the critical depinning force
at $f=1.0$ and $f=2.0$,
similar 
to what is observed for the depinning of vortices in
type-II superconductors with  periodic pinning \cite{7,9,10,11,12}. 
For this value of $F_{p}$ and
pinning site density,
these commensuration peaks are weak; however, as
$F_{p}$ decreases, the peaks become much 
more pronounced
since the commensurate depinning threshold is dominated by the
pinning strength and decreases linearly with $F_p$, but
the incommensurate depinning threshold is dominated by the
skyrmion-skyrmion interactions and decreases faster than
linearly with $F_p$.
A similar effect occurs
for superconducting vortex systems with periodic pining \cite{7,69,70}. 
Figure~\ref{fig:20} also shows that the PS phase appears
for
$f > 0.6$ and increases in extent with 
increasing
$f$.
The extent of phase II
deceases with increasing
$f$
since the
increase in the number of skyrmions flowing through
the interstitial regions causes
skyrmions at the pinning sites to be dislodged 
at lower drives,
triggering a transition to the phase III flow.

\begin{figure}
\includegraphics[width=\columnwidth]{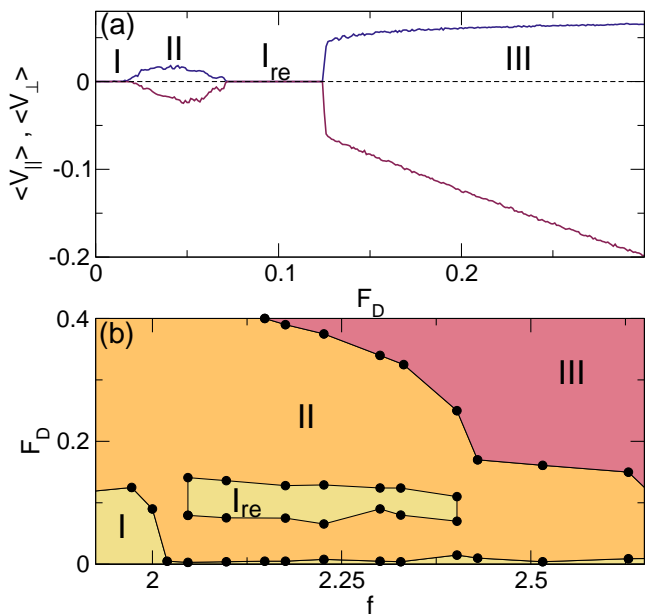}
  \caption{(a) $\langle V_{||}\rangle$ (blue) and
    $\langle V_{\perp}\rangle$ (red) at
    $f=2.33$ for the system in
    Fig.~\ref{fig:20} with $n_p=0.1975$, $\alpha_m/\alpha_d=9.96$,
    and $F_p=2.0$
    showing 
    a reentrant pinning effect, I$_{re}$.
    (b) A blowup of the dynamic phase diagram from
    Fig.~\ref{fig:20} as a function of $F_D$ vs $f$ showing the
    reentrant pinning which occurs for
    $2.04 \leq f < 2.4$.  
}
\label{fig:21}
\end{figure}

We observe a reentrant pinning regime in Fig.~\ref{fig:20}
for $ 2.04 \leq f \leq 2.4$,
where the skyrmions
enter phase II flow
but then form a new pinned configuration when the drive is
increased.
In Fig.~\ref{fig:21}(a) we plot both
$\langle V_{||}\rangle$ and
$\langle V_{\perp}\rangle$
versus $F_D$ at
$f=2.33$
showing
a transition from the initial pinned phase I
to 
the interstitial flow phase II, which is then followed by
a transition to a
reentrant pinned phase I$_{re}$ before the system
returns to phase II flow and finally
transitions into the disordered phase III flow.
The 
reentrant pinned phase
can be viewed as an example of a clogging effect.
In Fig.~\ref{fig:16}(d), the $f=3.0$ configuration is composed of
alternating interstitial dimers, which first
begin to form when $f>2.0$.
For $2.04 \leq f < 2.4$, a fraction of the system
contains interstitial dimers at $F_D=0$.
In this regime,
at the initial depinning transition
when the effect of the Magnus force is negligible,
the dimers align with the drive along a symmetry direction of the pinning lattice.
As the drive increases and the Magnus force begins to rotate the direction of
motion of the skyrmions, the dimers also rotate to follow the flow.  A dimer that has
rotated beyond a critical angle can become trapped if it abruptly
becomes oriented
perpendicular to the driving direction in the interstitial region between pinning
sites.  This ``dimer bridge'' blocks the flow locally, and if all the skyrmion dimers in the
system form such dimer bridges, the flow stops completely, producing the reentrant
pinned phase I$_{re}$.
When the drive is increased further,
some of the dimer bridges break apart and the system can reenter phase II flow.
At even higher drives, the skyrmions at the pinning sites depin
and disordered phase III flow appears. 
In Fig.~\ref{fig:21}(b) we show a blow up of the
dynamic phase diagram from Fig.~\ref{fig:20} as a function of $F_D$ versus $f$ over
the range
$1.9 < f < 3.0$, where we highlight phase I, II, and III flow.
Many of the phases
in
Fig.~\ref{fig:20}
are absent in the overdamped
limit,
where the PS phase does not occur and the MC phase is replaced by a moving smectic.
The reentrant pinning phenomena is also lost
in the overdamped limit since the dimers all remain aligned with the driving
direction,
whereas
a finite Magnus force
causes the dimers to move at an angle to the driving direction, making them
more likely 
to create a
clogged configuration.  

\section{Varied Pinning Density}

\begin{figure}
\includegraphics[width=\columnwidth]{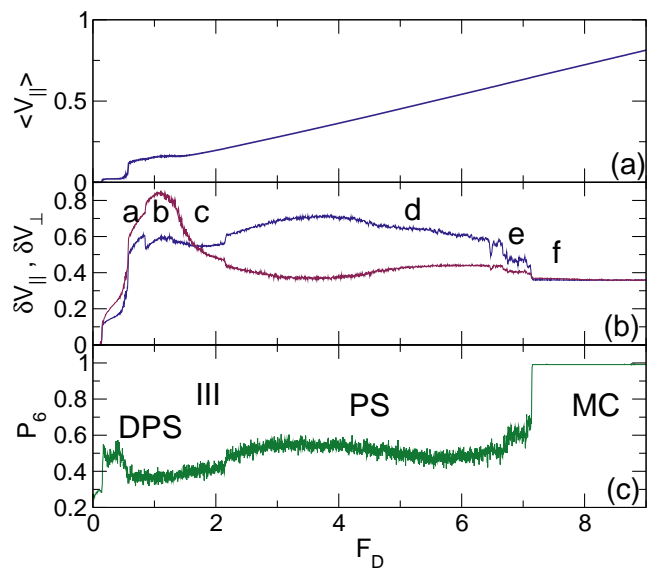}
\caption{
  (a) $\langle V_{||}\rangle$ vs $F_{D}$ for a system with
  $f=1.0117$, $F_p=2.0$, $\alpha_m/\alpha_d=9.96$, and
  pinning density $n_p=0.436$.
  (b) The corresponding $\delta V_{||}$ (blue) and $\delta V_{\perp}$ (red) vs $F_{D}$.
  The letters a to f indicate the values of $F_D$ at which the images in Fig.~\ref{fig:24}
  were obtained.
  (c) The corresponding $P_{6}$ vs $F_{D}$.
  DPS: diagonal phase separated; III: disordered flow; PS: phase separated; MC: moving
  crystal.
}
\label{fig:22}
\end{figure}

\begin{figure}
\includegraphics[width=\columnwidth]{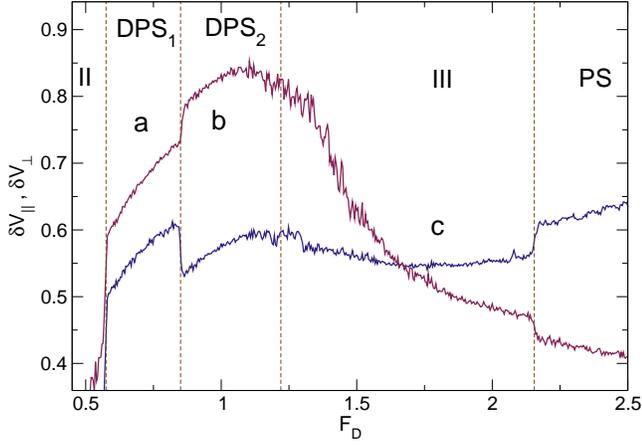}
\caption{ A blowup of the
  $\delta V_{||}$ (blue) and
  $\delta V_{\perp}$ (red) vs $F_{D}$ curves
  from Fig.~\ref{fig:22}(b) highlighting 
  the subphases DPS$_1$ and DPS$_2$ within the DPS phase.
  The subphases have different stripe orientations as illustrated
  in Fig.~\ref{fig:24}(a,b).
  The letters a, b, and c indicate the values of $F_D$ at which the
  images in Fig.~\ref{fig:24} were obtained.
  The vertical dashed lines are guides to the eye to 
  show the different phases.
  II: interstitial flow; DPS$_1$ and DPS$_2$: the two subphases of the diagonal
  phase separated state; III: disordered flow; PS: phase separated.
}
\label{fig:23}
\end{figure}

We next consider the effect of changing the pinning density $n_{p}$ while
fixing 
$f=1.0117$,
$F_{p} = 2.0$, and $\alpha_m/\alpha_d=9.96$.
In general, we find that the PS phase appears for sufficiently large $n_p$, and
that the DPS phase can also arise.
In Fig.~\ref{fig:22}
we plot $\langle V_{||}\rangle$,
$\delta V_{||}$, $\delta V_{\perp}$, and 
$P_{6}$ versus $F_{D}$ for a system with $n_{p} = 0.436$.
The pinned phase occurs for $F_{D} <  0.155$, while
phase II appears in the range
$0.155 \leq F_{D} < 0.575$.
In the DPS phase, which extends over the range
$0.575 \leq F_{D} <  1.35$, 
$\langle V_{||}\rangle$ is constant or slightly decreasing
with increasing $F_{D}$.
In phases III and PS, $\langle V_{||}\rangle$ increases with
increasing $F_D$.
At higher drives, the PS-MC transition
takes place
in a series of steps
that appear 
as a jump down in $\delta V_{||}$ and $\delta V_{\perp}$
along with a small increase in $P_{6}$ over the range
$6.4 < F_{D} < 7.1$,
followed by
a jump up in $P_{6}$ to  $P_6 \approx 1.0$
in the MC phase.

\begin{figure}
\includegraphics[width=\columnwidth]{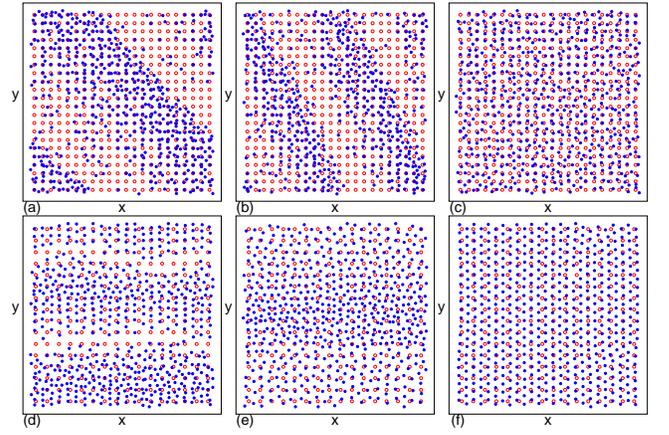}
\caption{ Skyrmion (blue dots) and pinning site (red circles)
  locations
  for the system in
  Fig.~\ref{fig:22} and Fig.~\ref{fig:23}
  with $f=1.0117$, $F_p=2.0$, $\alpha_m/\alpha_d=9.96$,
  and $n_p=0.436$ obtained at values of $F_D$ marked by
  the letters a to f in those figures.
  (a) At $F_{D} = 0.7$ in the DPS$_1$ subphase,
  the stripes are aligned at an angle of nearly
  $45^\circ$ with respect to the driving direction.
  (b) At $F_{D} = 1.0$ in the DPS$_2$ subphase,
  the stripes are aligned at an angle of nearly $67^\circ$ with respect
  to the driving direction.
  (c) Disordered phase III flow at $F_{D} = 1.8$.
  (d) At $F_{D} = 4.47$ in the PS phase,
  there are two moving stripes aligned with the driving direction.
  (e) At $F_{D} = 7.0$, close to the PS-MC transition, the two stripes
  begin to merge.
  (f) The MC phase at $F_{D} = 7.5$. 
}
\label{fig:24}
\end{figure}

As illustrated in Fig.~\ref{fig:23},
which shows a blowup of
$\delta V_{||}$ and $\delta V_{\perp}$ versus $F_D$
from Fig.~\ref{fig:22}(b)
over the range $0.45 < F_{D} < 2.5$,
there are two subphases in the DPS phase,
DPS$_1$ and DPS$_2$, that are distinguished by different stripe orientations.
The II-DPS$_1$ transition appears as
an upward jump in both $\delta V_{||}$ and $\delta V_{\perp}$,
while at the DPS$_1$-DPS$_2$ transition near
$F_D=0.85$,
there is a sharp jump down in $\delta V_{||}$ accompanied by
a jump up
in $\delta V_{\perp}$.
In Fig.~\ref{fig:24}(a), we show the DPS$_1$ subphase for the
$n_p=0.436$ system in Fig.~\ref{fig:22}
at
$F_{D} = 0.7$.
The skyrmions form dense bands at an angle of nearly $45^\circ$ with respect
to the driving direction.
At $F_{D} = 1.0$ in the DPS$_2$ subphase,
the bands are oriented at a steeper angle of
$67^{\circ}$ with respect to the
driving direction, as illustrated in Fig.~\ref{fig:24}(b).
The jumps in $\delta V_{||}$ and $\delta V_{\perp}$ near $F_{D} = 0.85$
in Fig.~\ref{fig:23} are associated with a reorientation of the stripes
at the DPS$_1$-DPS$_2$ transition.
After the reorientation occurs, 
the stripes are aligned
closer to the perpendicular
direction,
causing $\delta V_{\perp}$ to jump up while
$\delta V_{||}$ jumps down, as shown in
Fig.~\ref{fig:23} near $F_{D} = 0.84$. 
Near $F_{D} = 1.22$, the stripes start to break up, and the system enters
a uniform density disordered flow phase III,
illustrated in Fig.~\ref{fig:24}(c) at
$F_{D} = 1.8$.
In the PS phase,
a stripe aligned with the driving direction appears, 
similar to the state illustrated earlier;
however, as $F_D$ increases,
the stripe breaks into two parallel stripes,
as shown in Fig.~\ref{fig:24}(d) at $F_D=4.47$.
For  $6.7 < F_{D} < 7.1$, the two stripes broaden
and merge into a single
density modulation extending across the entire sample, 
as shown at $F_D=7.0$ in Fig.~\ref{fig:24}(e),
and finally the density evens out and the MC phase
emerges, as illustrated in Fig.~\ref{fig:24}(f)
at $F_{D} = 7.5$.

\begin{figure}
\includegraphics[width=\columnwidth]{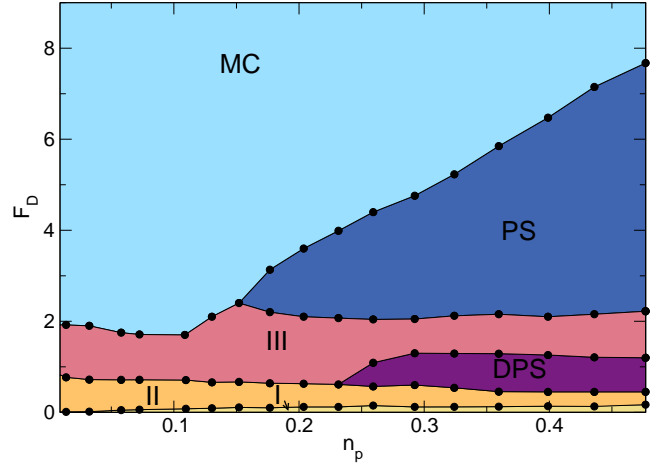}
\caption{Dynamic phase diagram as a function of $F_D$ vs $n_{p}$
  for the system in Fig.~\ref{fig:22}
  with $f=1.0117$,  $F_{p} = 2.0$, and $\alpha_m/\alpha_d=9.96$.
  I: pinned; II: interstitial flow; III: disordered flow; DPS: diagonal phase separated;
  PS: phase separated; MC: moving crystal.
}
\label{fig:25}
\end{figure}

In Fig.~\ref{fig:25} we plot the dynamic phase
diagram as a function of $F_{D}$ versus $n_{p}$ for the system 
in Fig.~\ref{fig:22} with
$f=1.0117$, $F_p=2.0$, and $\alpha_m/\alpha_d=9.96$.
The pinned phase I disappears
at small $n_{p}$ since
the distance between the directly pinned skyrmions
increases as the pinning density decreases,
and therefore the strength of 
the caging potential experienced by
the interstitial skyrmions due to the skyrmion-skyrmion
interactions also decreases until the interstitial skyrmions
can no longer be trapped.
The PS phase only occurs for $n_{p} > 1.7$ and
grows in extent with increasing $n_{p}$,
similar to the results for varied skyrmion density, where the PS
phase emerges only for sufficiently high skyrmion density,
indicating that the PS phase results from collective skyrmion interactions.
The DPS phase appears when $n > 0.235$, and there
is an additional line within the DPS phase
(not shown) separating the two different orientations DPS$_1$ and DPS$_2$
illustrated in Fig.~\ref{fig:24}(a,b). 
The III-PS transition falls at a nearly constant value of $F_D=F_p=2.0$,
indicating that the PS phase forms once $F_D>F_p$ and all the skyrmions depin.

We observe similar results for skyrmions interacting
with triangular pinning arrays.
In general,
the PS phases are more pronounced
and exhibit shaper transitions
when the pinning is in a periodic array rather than randomly placed;
however, the PS phase is a robust feature of skyrmions in systems with strong pinning
and strong Magnus force, regardless of the pinning arrangement.

\section{Summary} 
We have examined the nonequilibrium phases of
skyrmions interacting with periodic arrays of pinning sites 
using a particle-based model for the skyrmions
in which the dynamics
is governed  by both damping and Magnus terms.
As a function of system parameters,
we find a rich variety of distinct nonequilibrium phases
and transitions between these phases
which produce clear signatures in the velocity-force curves,
fluctuations,
and skyrmion ordering.
There are pronounced differences between these phases 
and the collective nonequilibrium  phases 
found for overdamped particles moving over
periodic substrates, such as vortices in type-II superconductors
or colloids on optical trap arrays.
The Magnus force causes the skyrmions to
move at an angle with respect to the external drive;
however, when pinning is present
this skyrmion Hall angle becomes drive dependent.
As a result, when the pinning is sufficiently strong,
the velocity can become spatially heterogeneous,
with skyrmions in different portions of the sample experiencing
different effective skyrmion Hall angles,
producing an instability that leads to the formation of a clustered or
phase separated state.
When the drive is high enough, the effectiveness of the pinning diminishes 
and the skyrmions enter a
uniform moving crystal phase. 
In addition to the phase separated 
state
we find
a pinned phase,
an interstitial flow phase in which pinned and interstitial
skyrmions coexist,
a disordered flow phase, and 
a moving crystal state.  
When the pinning is weak or the Magnus force is small,
the phase
separated state is lost.
The segregated phase we observe is very similar to that
found in recent continuum-based simulations
of skyrmions in the strong substrate limit,
indicating that the
dynamical clustering effects
can be captured without
taking the internal  modes of the skyrmions into account.  
We also find
distinct types of moving phase separated states
when the stripe-like clusters form at
different orientations with respect to the pinning lattice symmetry.  
For deceasing pinning strength,
we
observe a transition from a commensurate  or partially
commensurate state to
a floating solid, which is similar to the Aubry transition found in
colloidal systems.
For stronger pinning
there are a series of commensurate-incommensurate
transitions where different types of skyrmion
crystalline states can be stabilized as a function of the ratio of the number of
skyrmions to the number of pinning sites.
These states are the same as those observed in an
overdamped system since the Magnus force does
not change the pinned ground state 
configurations.

\begin{acknowledgments}
We gratefully acknowledge the support of the U.S. Department of
Energy through the LANL/LDRD program for this work.
This work was carried out under the auspices of the 
NNSA of the 
U.S. DoE
at 
LANL
under Contract No.
DE-AC52-06NA25396 and through the LANL/LDRD program.
\end{acknowledgments}

\end{document}